\documentclass[aps,prd,superscriptaddress,nofootinbib,tighten,preprint]{revtex4}
\usepackage{color}
\usepackage{latexsym}
\usepackage{amssymb}
\usepackage{amsmath}
\usepackage{graphicx}
\usepackage{hyperref}
\begin{document}

\title{Sensitivity to neutrino decay with atmospheric neutrinos at INO}
%-----------------------------------------------------------------------------------%
\author{Sandhya Choubey}
\email{sandhya@hri.res.in}
\affiliation{Harish-Chandra Research Institute, Chhatnag Road, Jhunsi, Allahabad 211 019, India}
\affiliation{Department of Physics, School of
Engineering Sciences, KTH Royal Institute of Technology, AlbaNova
University Center, 106 91 Stockholm, Sweden}
\affiliation{Homi Bhabha National Institute, Training School Complex, Anushakti Nagar,
     Mumbai 400085, India} 

\author{Srubabati~Goswami}
\email{sruba@prl.res.in}
\affiliation{Physical Research Laboratory, Navrangpura, Ahmedabad 380 009, India}

\author{Chandan Gupta} 
\email{chandan@prl.res.in}
\affiliation{Homi Bhabha National Institute, Training School Complex, Anushakti Nagar,
     Mumbai 400085, India} 
\affiliation{Physical Research Laboratory, Navrangpura, Ahmedabad 380 009, India}
\author{S.~M.~Lakshmi}
\email{slakshmi@prl.res.in}
\affiliation{Physical Research Laboratory, Navrangpura, Ahmedabad 380 009, India}

\author{Tarak Thakore}
\email{tarakstar@gmail.com}
\affiliation{Louisiana State University, Baton Rouge 70803, LA}

\date{\today} %Dec 18, 2015
\bigskip

\begin{abstract}
Sensitivity of the magnetised Iron CALorimeter (ICAL) detector at the proposed India-based Neutrino Observatory (INO) to invisible decay 
of the mass eigenstate $\nu_3$ using atmospheric neutrinos is explored. A full three-generation analysis including earth matter 
effects is performed in a framework with both decay and oscillations. The wide energy range and baselines offered by atmospheric
neutrinos are shown to be excellent for constraining the $\nu_3$ lifetime. We find that with an exposure of 500 kton-yr the ICAL
atmospheric experiment could constrain the $\nu_3$ lifetime to $\tau_3/m_3>1.51\times10^{-10}$~s/eV at the 90\% C.L. This is two 
orders of magnitude tighter than the bound from MINOS. The effect of invisible decay on the precision measurement of $\theta_{23}$ 
and $|\Delta{m^2_{32}}|$ is also studied. 
\end{abstract}

\maketitle

\section{Introduction}
Neutrino oscillation experiments spanning various 
energy ranges and baselines have helped in establishing the fact that neutrinos oscillate from one flavor to another. Most of the neutrino oscillation parameters have been pinned down and are now 
known rather precisely. \footnote{See \cite{nufit-2017,pdg-nkmu-petc} for the current global best-fit values of the oscillation parameters 
and the references therein to all the past and on-going experimental efforts.} 
The main open questions remaining in neutrino oscillation physics are neutrino mass hierarchy, octant of the  mixing 
angle $\theta_{23}$ and the value of the CP phase $\delta_{CP}$. Several experiments are running or 
are being planned in order to answer the above-mentioned questions. The leading experimental proposals for the 
future include the long-baseline experiments DUNE \cite{dune} and T2HK \cite{t2hk}, reactor experiments 
JUNO \cite{juno} and RENO50 \cite{reno50-1}, and atmospheric neutrino experiments PINGU \cite{pingu}, ORCA \cite{orca}
and ICAL \cite{WP,ICAL}. It is expected that the neutrino oscillation probabilities would change in the presence 
of new physics. This could be used to constrain new physics scenarios at neutrino oscillation experiments. At the 
same time, a given new physics scenario could also interfere with the measurement of the standard neutrino 
oscillation parameters and hence pose a challenge to the proposed experiments, unless ways are found to cancel out
their effects through synergistic measurements at multiple experiments. One such new physics scenario is the decay
of neutrino during its flight from the source to the detector.  

While there is no observational evidence in support for unstable neutrinos, since they are massive, its not unlikely
that they would decay. Radiative decays of neutrinos are very severely constrained by cosmological data. Since the 
measured neutrino masses suggest that the neutrinos would radiatively decay in the microwave energy range, the most
stringent bounds are provided by cosmic microwave background data \cite{mirizzi2007}, making radiative decay of
neutrinos totally uninteresting for neutrino oscillation experiments. However, there still remains the possibility 
that neutrinos could decay into a lighter fermion state and a beyond standard model boson. The Majoron 
model \cite{Chikashige:1980ui,Gelmini:1980re,Gelmini:1983ea} for instance allows the following decay modes 
for Majorana neutrinos: $\nu_i \to \nu_j + J$ or $\nu_i \to \bar\nu_j + J$, where $\nu_j$ and $\bar\nu_j$ are 
lighter neutrino and anti-neutrino states and $J$ is a Majoron. The Majoron in principle could belong to either a 
singlet or a triplet representation of the standard model gauge group. But 
the triplet model is severely constrained \cite{Gelmini:1980re,Gelmini:1983ea}and hence $J$ must predominantly be an electroweak singlet.
If the final state fermion is a lighter active neutrino, the decay is called {\it visible decay}. On the other hand, if the 
final state fermion is a sterile state with no standard model interaction, then the decay scenario is termed 
{\it invisible decay}. Even for Dirac neutrinos in extensions of the standard models one could write down terms in 
the Lagrangian coupling neutrinos with a light scalar boson and light right-handed neutrinos allowing the decay mode
$\nu_i \to \bar\nu_{iR} + \chi$, where $\bar\nu_{iR}$ is a right-handed singlet neutrino and $\chi$ is an 
iso-singlet scalar carrying lepton number $+2$ \cite{Acker:1991ej,Acker:1992eh}. In this paper, we will work in a scenario where the 
final state particles remain invisible to the detector. 

The lifetime of $\nu_2$ (and $\nu_1$) is constrained by the solar neutrino experiments. Neutrino decay as a solution
to the solar neutrino deficit problem was suggested in \cite{Bahcall:1972my}, however, now we know that neutrino 
decay alone cannot explain this deficit. Attempts to constrain the neutrino lifetime by considering neutrino decay
as a subdominant effect along with the leading LMA-MSW solution was done 
in \cite{Acker:1993sz,Berezhiani:1991vk,Berezhiani:1992xg,Choubey:2000an,Bandyopadhyay:2001ct,Joshipura:2002fb,
Bandyopadhyay:2002qg,Berryman:2014qha,Picoreti:2015ika}. Most of these studies considered the invisible decay 
scenario. Since $U_{e3}$ is small, the $\nu_e$ state mostly resides in the $\nu_2$ and $\nu_1$ states and hence 
all of these studies worked in the two-generation framework. Bounds on the lifetime of $\nu_2$ was obtained from 
a global analysis of solar neutrino data in \cite{Bandyopadhyay:2002qg} where the impact of the Sudbury Neutrino 
Observatory neutral current data was highlighted. It was shown that the bound on $\nu_2$ lifetime was 
$\tau_2/m_2 > 8.7\times 10^{-5}$~s/eV at 99\% C.L. for a 3 parameter fit. This bound was revisited in 
\cite{Berryman:2014qha} (see also \cite{Picoreti:2015ika}) where the authors obtained the 95\% C.L. limit 
$\tau_2/m_2 > 7 \times 10^{-4}$~s/eV for both normal and inverted mass hierarchy and 
$\tau_1/m_1 > 4\times 10^{-3}$~s/eV for inverted mass hierarchy. These results are very consistent with the 
earlier analysis of \cite{Bandyopadhyay:2002qg} where the 95\% C.L. limit for a one parameter fit is seen to 
be $\tau_2/m_2 > 4.4\times 10^{-4}$~s/eV. The corresponding constraints from SN1987A are 
stronger \cite{Frieman:1987as}. 

Limits on the lifetime of $\nu_3$ come from the atmospheric and long-baseline neutrino experiments. Like in the
case of solar neutrinos, any fit with neutrino decay alone \cite{LoSecco:1998cd,Lipari:1999vh} is unable to explain 
the atmospheric neutrino zenith angle data. A lot of work has gone into considering decay along with oscillations. 
The analyses can be broadly classified into two categories depending on the model used. If one considers decay 
of $\nu_3$ to a state with which it oscillates, then the bounds coming from K-decays \cite{Barger:1981vd} restrict the corresponding 
mass squared difference between them to $\Delta m^2 > 0.1$~eV$^2$ \cite{Barger:1998xk}. 
However, if the state to which $\nu_3$ decays 
is a sterile state then the $\Delta m^2$ driving the leading oscillations of $\nu_\mu$ is unconstrained. The former 
case is that of decay to active neutrinos and was studied in the context of atmospheric neutrinos 
in \cite{Barger:1998xk,Fogli:1999qt} and no good fit was found. The latter is the invisible decay scenario to sterile neutrinos 
and was 
analysed against the atmospheric neutrino data in \cite{Choubey:1999ir,Barger:1999bg,Ashie:2004mr,ggm-atms-nu}. The
invisible decay case can be again classified into two. In one case we can make the assumption that 
$\Delta m^2 \ll 10^{-4}$~eV$^2$, causing it to drop out of the oscillation probability. The authors
of \cite{Barger:1999bg} argued that this could explain the Super-Kamiokande atmospheric neutrino data, however,
the Super-Kamiokande collaboration itself reported \cite{Ashie:2004mr} that this scenario was not supported by 
their data. The other case of invisible decay is when $\Delta m^2$ is left free in the fit to be determined by 
the data. This case was first proposed by some of us in \cite{Choubey:1999ir}. The results of \cite{Choubey:1999ir}
were updated in \cite{ggm-atms-nu} where the authors obtained the limit $\tau_3/m_3 > 2.9 \times 10^{-10}$ s/eV for invisible decay at 
the 90\% C.L. from a combined analysis of Super-Kamiokande atmospheric and MINOS data. More recently, the analysis 
of oscillation plus invisible decay scenario with unconstrained $\Delta m^2$ was performed in \cite{gomes} in the context of 
MINOS and T2K data and gave a bound $\tau_3/m_3 > 2.8 \times 10^{-12}$~s/eV at 90\% C.L. 
The constraint for the visible decay scenario using the MINOS and T2K charged as well as neutral current data was 
performed in \cite{Gago:2017zzy}.  
The bounds on neutrino 
lifetime could be improved considerably by observations at IceCube using cosmological baselines
\cite{Beacom:2002vi,Maltoni:2008jr,Pagliaroli:2015rca,Bustamante:2016ciw}. 

All the above mentioned papers which considered neutrino decay alongside oscillations performed their analysis in 
the framework of two-generations and did not take earth matter effects into account. Recently 
a three-generation analysis including earth matter effect and decay in the context of the Deep Underground 
Neutrino Experiment (DUNE) was performed in \cite{coloma-decay}  for visible decays and  \cite{pramanik} for 
invisible decays. It was shown that DUNE could improve the bound on $\tau_3/m_3$ for the invisible decay case by at least an 
order of magnitude compared to the current limits from MINOS and T2K. In this work, we consider invisible neutrino 
decay within a three-generation oscillation framework in the context of atmospheric neutrinos and include earth matter
effects. Atmospheric neutrinos span many orders of magnitude in energy and baseline. Since the effect of
neutrino decay increases for lower energies and longer baselines, atmospheric neutrino experiments are expected to give 
a tighter bound on $\tau_3/m_3$ than the proposed long-baseline experiments. We will study the sensitivity of the
atmospheric neutrinos at INO to neutrino decay. 

The India-based Neutrino Observatory (INO) is a proposed underground laboratory in India, which plans to house a 50
kton magnetised Iron CALorimeter (ICAL). The detector will be mainly sensitive to muon type neutrinos, which are 
detected through the observation of a muon track and the accompanying  hadron shower in a charged current interaction. 
The detector response to muons \cite{mupaper1,mthesis,peripheralmu,kthesis} and hadrons \cite{hres1,hrest,lsmthesis,mmdthesis}
have been performed via the Geant4-based \cite{geant1,geant2,geant3} detector simulation code for ICAL. This detector owing to its magnetisation can distinguish between 
neutrino and anti-neutrino events which makes it an excellent detector to determine the neutrino mass 
hierarchy \cite{WP,TAhie,Ghosh:2013mga,3dMMD,hi-mu,Ali-hie}. ICAL will also perform precision measurements of $|\Delta{m^2_{32}}|$ and the 
mixing angle $\theta_{23}$ \cite{WP,3dMMD,hi-mu,TApre, DJ-nu, DJ-nuanu, monojet}. In addition, there are a variety of new physics scenarios which 
could be constrained and/or discovered at ICAL. Some of the new physics scenarios studied by the INO collaboration 
include, CPT violation \cite{Ani-CPT}, dark matter \cite{nitali-DM}, non-standard neutrino interactions 
\cite{Choubey:2015xha} and sterile neutrino oscillations \cite{shibu-sterile}. In this work we will study in detail the 
sensitivity of ICAL to invisible neutrino decay using the full physics analysis simulation framework of ICAL. We will also study the
effect of invisible neutrino decay on the precision measurement of $|\Delta{m^2_{32}}|$ and the mixing angle $\theta_{23}$. 

The paper is organised as follows. The scenario of invisible decay plus oscillations for three--generation 
mixing and oscillations in earth matter are discussed in Section~\ref{decay-prob}. The simulation of events 
and $\chi^2$ analysis are explained in Section~\ref{nu-evt-gen}. In Section~\ref{a3-results} we present our 
results for the sensitivity to the decay parameter $\tau_3/m_3$. The effects of the presence of decay on the 
precision measurements of $\sin^2\theta_{23}$ and $|\Delta{m^2_{32}}|$ are discussed in Sections~\ref{pre-tt23-wd} and \ref{pre-dm232-wd}
respectively. The exclusion contours are presented in Section~\ref{contours}. Conclusions are presented in 
Section~\ref{conclusion}.
  
\section{Invisible decay and oscillations in the presence of matter}\label{decay-prob}
 In this section we consider the oscillations and decay of $\nu_3$ in the 
 presence of matter. Let the state $\nu_3$ decay invisibly via $\nu_3\rightarrow \nu_s+J$, where
 $J$ is a pseudo-scalar and $\nu_s$ is a sterile neutrino. Since $\nu_s$ does not mix with the 
 three active neutrinos, the mixing matrix $U$ in vacuum \cite{upmns1,upmns2,upmns3} is given by :
\begin{equation}
 U = \begin{pmatrix}
c_{12}c_{13} & s_{12}c_{13} & s_{13}e^{-i\delta} \\
-c_{23}s_{12}-s_{23}s_{13}c_{12}e^{i\delta} & c_{23}c_{12}-s_{23}s_{13}s_{12}e^{i\delta} & s_{23}c_{13} \\
s_{23}s_{12}-c_{23}s_{13}c_{12}e^{i\delta} & -s_{23}c_{12}-c_{23}s_{13}s_{12}e^{i\delta} & c_{23}c_{13}
\end{pmatrix},  
\label{upmns}
\end{equation}
where $c_{ij}~=~\cos\theta_{ij}$, $s_{ij}~=~\sin\theta_{ij}$; $\theta_{ij}$ are the mixing angles and $\delta$ is the CP violating phase. 

The mass of $\nu_s$ is such that when the hierarchy is normal, $m_s<m_1<m_2<m_3$. 
Since $\nu_s$ does not mix with the active neutrinos, the propagation equation is not affected by this. 
The effect of decay is included in the three-flavor evolution equation in the presence of earth matter as follows :
 \begin{equation}
  i\frac{d\tilde{\nu}}{dt} = \frac{1}{2E}\left[U\mathbb{M}^2U^\dagger + \mathbb{A}_{CC}\right]\tilde{\nu},
  \label{nuevematter}
  \end{equation}
  \begin{equation}
  \mathbb{M}^2~=~
  \begin{pmatrix}
  0 & 0 & 0 \\ 
  0 & \Delta{m^2_{21}} & 0 \\ 
  0 & 0 & \Delta{m^2_{31}}-i\alpha_3 
  \end{pmatrix}
  \,,~~{\rm and}~~
    \mathbb{A}_{CC}~=~
  \begin{pmatrix}
  A_{cc} & 0 & 0 \\ 
  0 & 0 & 0 \\ 
  0 & 0 & 0 
  \end{pmatrix}
  ,
 \end{equation}
where $E$ is the neutrino energy, $\alpha_3=m_3/\tau_3$ is the decay constant in units of~eV$^2$,
$m_3$ is the mass of $\nu_3$ and $\tau_3$ its rest frame life time. Since the term $\alpha_3$ appears in the propagation
equation along with $\Delta{m^2_{31}}$, it has to be in units of~eV$^2$. The conversion factor used here is 
$1~ \hbox{eV/s} = 6.58\times10^{-16}~ \hbox{eV}^2$.
The matter potential is 
\begin{equation}
A_{cc}=2\sqrt{2}G_Fn_eE=7.63\times10^{-5}\hbox{eV}^2~\rho(\hbox{gm/cc})~E(\hbox{GeV}),
\label{matt-pot}
\end{equation}
where, $G_F$ is the Fermi constant and $n_e$ is the electron number density in matter and $\rho$ is the matter 
density. For anti-neutrinos, both the sign of $A_{cc}$ and the phase $\delta$ in Eq.~(\ref{nuevematter}) are
reversed. 

\subsection{Effect of the decay term}\label{exp-aLbE}
\begin{figure}[htp]
  \centering
   \includegraphics[width=0.42\textwidth,height=0.4\textwidth]{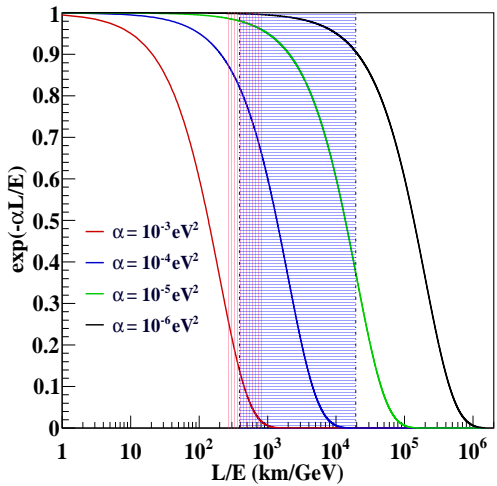}
   \caption{The value of $\exp{(-\alpha L/E)}$ as a function of $L/E$ for different values of
   the decay parameter $\alpha$. The red shaded region denotes the $L/E$ range accessible with 
   NO$\nu$A narrow band neutrino beam ($E=$ 1--3~GeV) the dashed blue shaded region indicates the
   range for $L$ = 9700~km, when $E$ is in the range 
   0.5--25~GeV.}
   \label{exp-aLbE-all}
  \end{figure}

The decay term is of  the form of 
$\exp{\left(-\alpha L/E\right)}$. 
No decay corresponds to $\alpha =0$ and the exponential term as 1 
whereas complete decay will be when the exponential term tends to 0. 
The effect of the decay parameter $\alpha$ for various $L/E$ values can be
understood from Fig.~\ref{exp-aLbE-all} in which 
$\exp{\left(-\alpha L/E\right)}$ vs $L/E$ is plotted for the values 
$\alpha=10^{-3}$, $10^{-4}$, $10^{-5}$ and $10^{-6}$~eV$^2$. 
%The following conclusions can be drawn from the figure. 
This figure gives an indication towards what are the values of 
$\alpha$ to which a given experiment spanning a specified $L/E$ range 
can be sensitive to. For instance the red shaded region in 
% and the blue shaded regions in
Fig.~\ref{exp-aLbE-all} indicates the $L/E$ range covered by the narrow band 
NO$\nu$A neutrino beam ($E$ = 1--3~GeV). 
It can be seen from the figure that NO$\nu$A's sensitivity is limited to 
larger values of $\alpha$; i.e $10^{-3}$ and 
$10^{-4}$~eV$^2$ for which the exponential terms shows substantial departure 
from the no decay value of 1. 
The blue shaded region corresponds to the
baseline $L$ = 9700 km with $E$ = 0.5--25~GeV, respectively.  
These are the typical values for an atmospheric neutrino experiment. 
This range of $L/E$ is sensitive to a wider range 
of $\alpha$ from $\sim$  $10^{-6}-10^{-3}$~eV$^2$ owing to the fact that 
it covers more $L/E$.  

%%%%%%%%%%%%%%%%%%%%%%%%%%%%%%%%%%%%%%%%%%%%%%%%%%%%%%%%%%%%%%%%%%%%%
%%%%%%%%%%%%%%%%%%%%%%%%%%%%%%%%%%%%%%%%%%%%%%%%%%%%%%%%%%%%%%%%%%%%%%%

 The ranges of $\exp{(-\alpha L/E)}$ values for various values 
 of $\alpha$ accessible for the specified range of $L/E$ for a given baseline is 
 shown in Table~\ref{eaLbE-range}.

%%%%%%%%%%%%%%%%%%%%%%%%%%%%%%%%%%%%%%%%%%%%%%%%%%%%%%%%%%%%%

  \begin{table}[htp]
  \centering
   \begin{tabular}{|c|c|c|c|c|c|}
   \hline
   $L$ (km) & $L/E$ (min)  & $L/E$ (max) & $\alpha$ (eV$^2$) & $\exp{(-\alpha L/E)}$ & $\exp{(-\alpha L/E)}$  \\
	    & (km/GeV) 	   & (km/GeV)   &  	            & (min)                 &  (max) \\
   \hline
       &			&		&	$10^{-3}$	&	0.016	&	0.254	\\
   810 &		270	&	810	& 	$10^{-4}$	&	0.663 	&	0.872	\\
       &			&		&	$10^{-5}$	&	0.959	& 	0.986	\\
       &			&		&	$10^{-6}$	&	0.996	& 	0.998	\\
       \hline
       &			&		&	$10^{-3}$	&	0	&	0.14	\\
  9700 &		388	&	19400	& 	$10^{-4}$	&	0 	&	0.82	\\
       &			&		&	$10^{-5}$	&	0.37	& 	0.98	\\
       &			&		&	$10^{-6}$	&	0.91	& 	1	\\
       \hline
  \end{tabular}
  \caption{Allowed ranges of $L/E$ in km/GeV for two fixed baselines 810 km 
  and 9700 km with detectable neutrino energies as 1--3 GeV and 0.5--25 GeV respectively. 
  The maximum and minimum values of $\exp{(-\alpha L/E)}$ for various $\alpha$ values for these 
  $L/E$s are also shown.}
  \label{eaLbE-range}
  \end{table}
%%%%%%%%%%%%%%%%%%%%%%%%%%%%%%%%%%%%%%%%%%%%%%%%%%%%%%%%%%%%%%%%%
For a given $L$, a broader range of $E$ will improve the sensitivity to
$\alpha$; on the other hand for a given $E$ range the sensitivity to $\alpha$ 
will increase if longer baselines are available. In principle any experiment which spans over a 
wide range of $L/E$ will have a better sensitivity to decay; with larger $L/E$s being sensitive to 
smaller values of $\alpha$ and vice versa. Atmospheric neutrino oscillation experiments fulfill this
exact requirement. If we consider the neutrino energy range of 0.5--25~GeV, atmospheric neutrinos will
span the $L/E$ range of [0.6, 25484]~(km/GeV) which includes all possible baselines from 15 km to the earth's diameter.
The INO ICAL detector becomes relevant in this context. Since ICAL can detect neutrinos in the range 0.5--25~GeV 
\cite{hi-mu} and since it is an atmospheric neutrino experiment, it will be sensitive to a wide range of 
 $\alpha$ values. As seen from Fig.~\ref{exp-aLbE-all} ICAL should give a sensitivity to $\alpha=10^{-6}$~eV$^2$
also. The sensitivity to low $\alpha$ values come from the low energy part of the spectrum, while the 
higher energy parts of the spectrum will help us rule out larger values of $\alpha$.  

%If ICAL can be optimised to detect energies less than 0.5 GeV, it will improve the sensitivity to smaller $\alpha$. 
%This is all the more important since the atmospheric neutrino spectrum peaks at 
%  lower energies and being able to access those energies will further improve the sensitivity to decay as well 
%  as other standard oscillation parameters. 
  
\subsection{Full three-flavor oscillations with decay in earth matter}
\label{prem-mod}

\begin{figure}
\centering
\includegraphics[width=0.42\textwidth,height=0.38\textwidth]{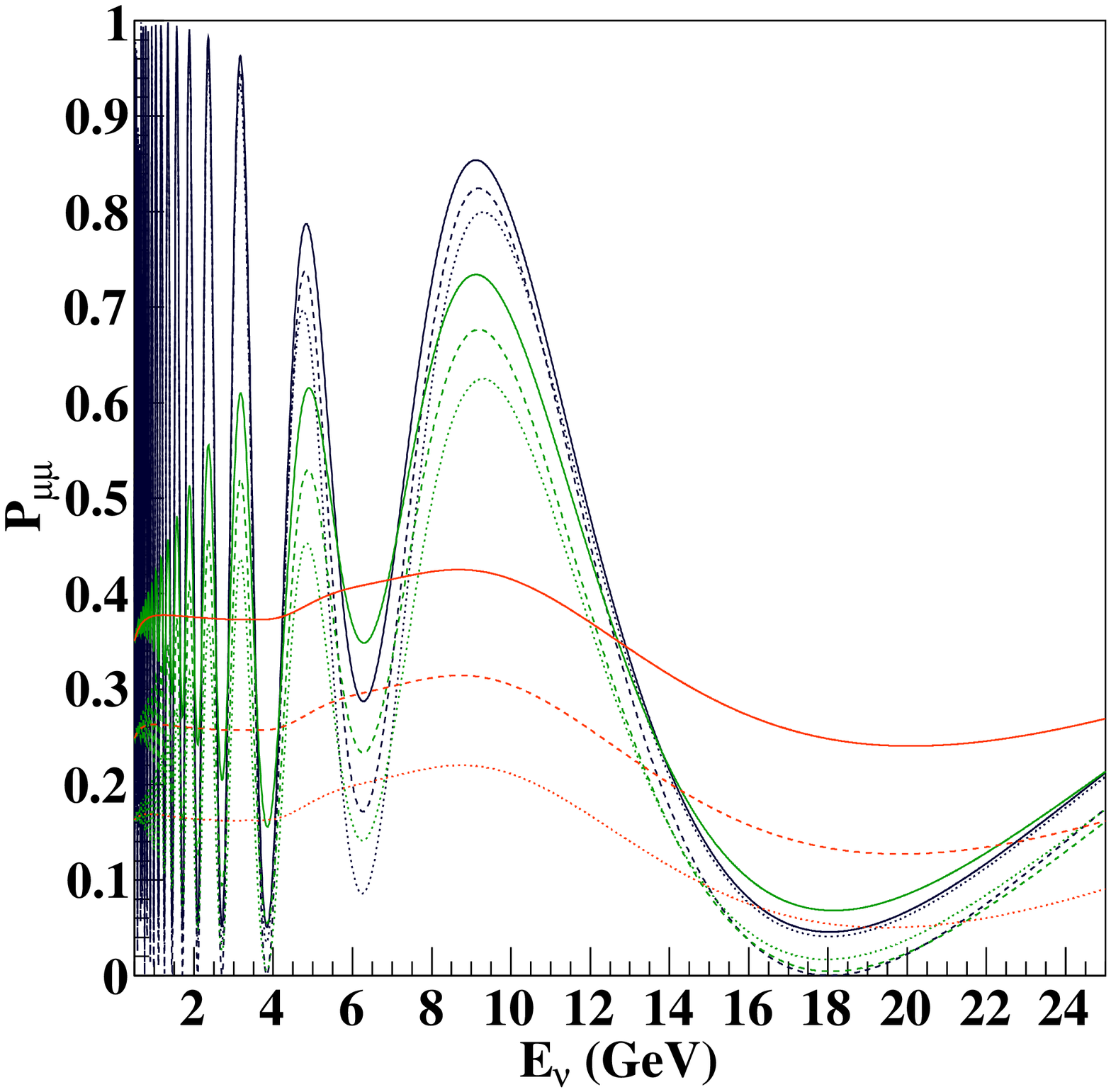} \includegraphics[width=0.42\textwidth,height=0.38\textwidth]{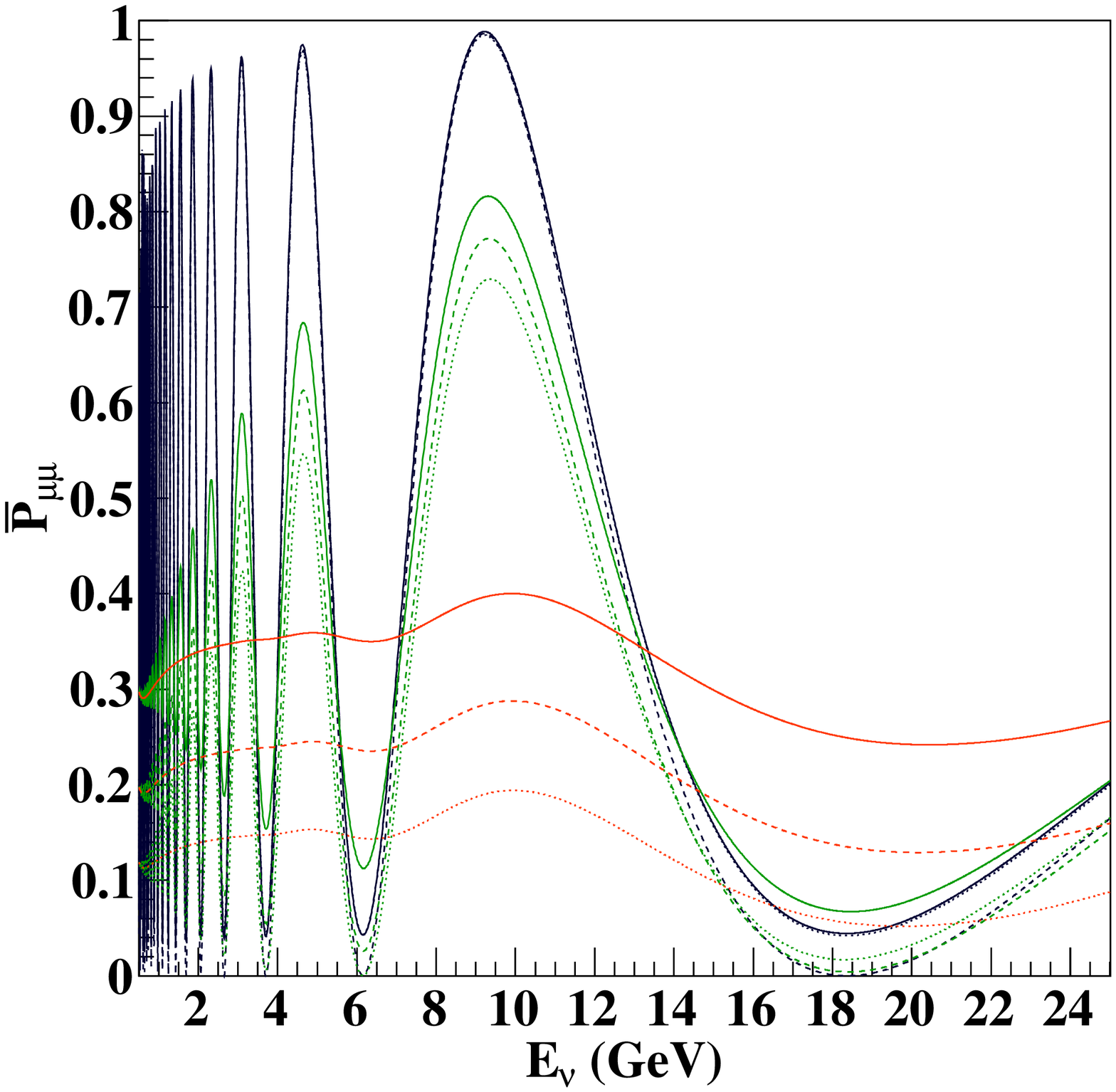}\\
\includegraphics[width=0.42\textwidth,height=0.38\textwidth]{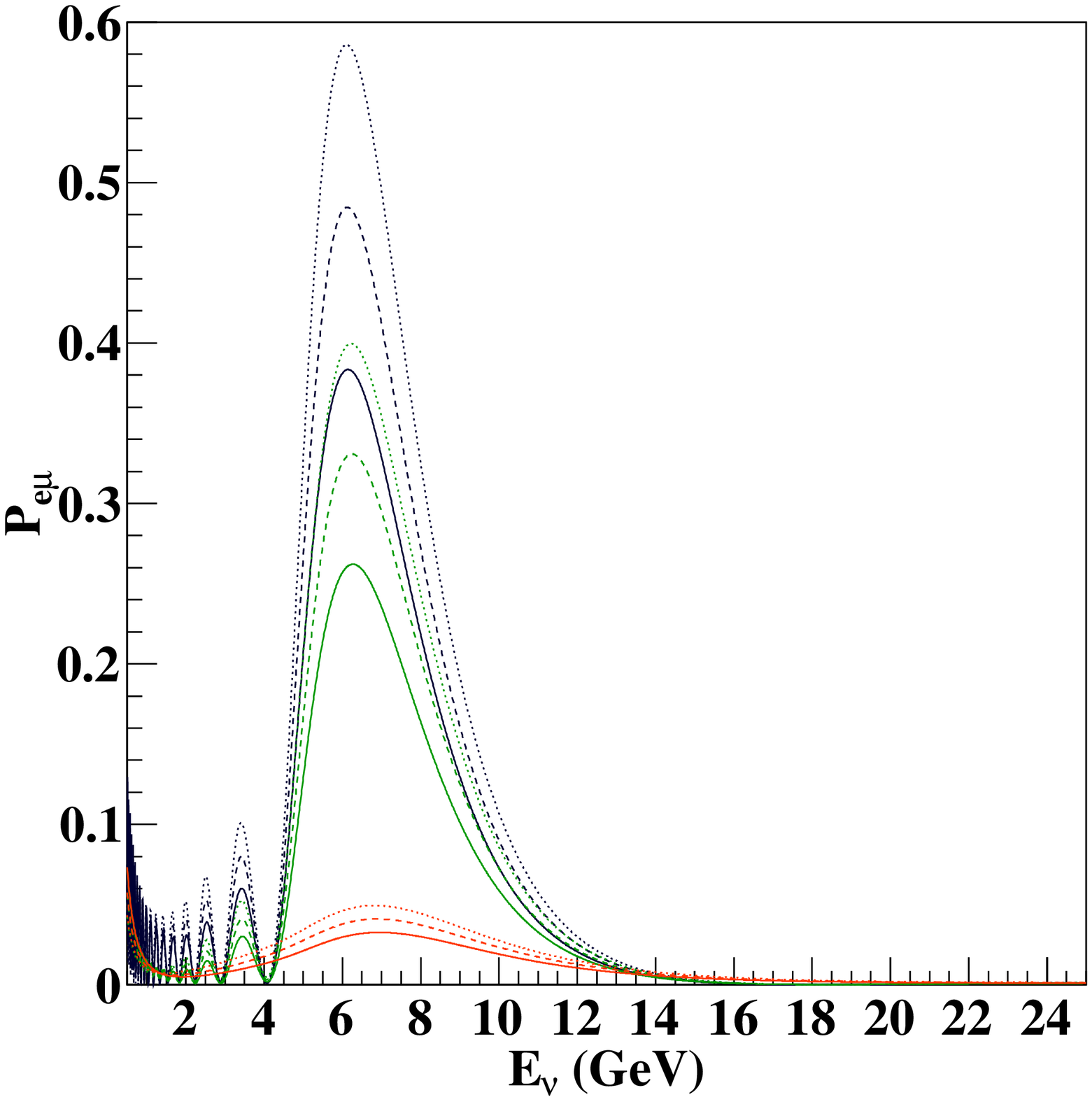} \includegraphics[width=0.42\textwidth,height=0.38\textwidth]{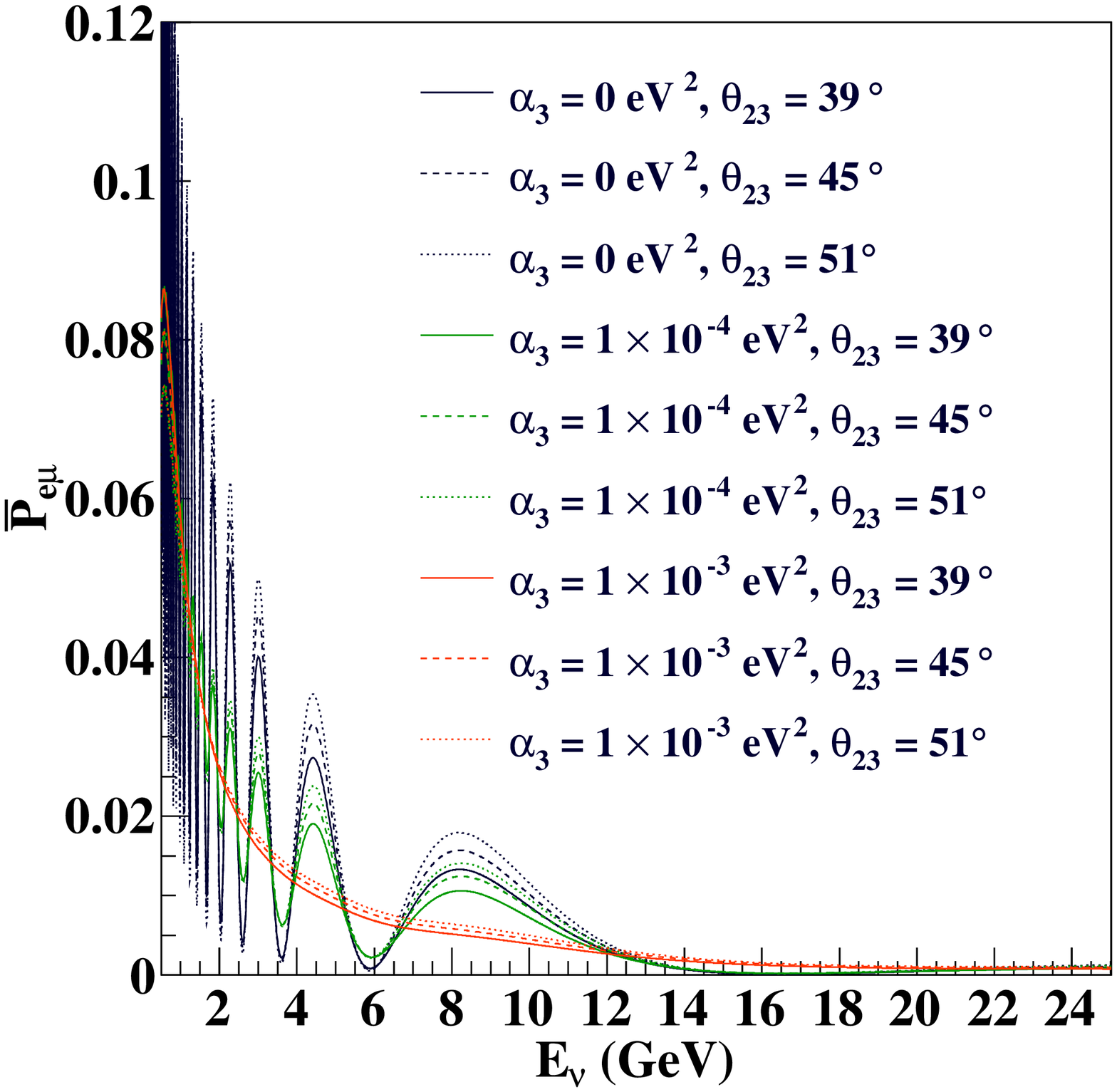}
\caption{Oscillation probabilities in matter for $\alpha_3=0,~1\times10^{-4}$ and $1\times10^{-3}$~eV$^2$ and 
$\theta_{23}=39^\circ,45^\circ$ and $51^\circ$, for the baseline $L$ = 9700~km in the energy 
range $E_\nu$ = 0.5--25~GeV. (Top-left) $P_{\mu\mu}$ and (top-right) $\bar{P}_{\mu\mu}$; (bottom-left) $P_{e\mu}$ and 
(bottom-right) $P_{\mu\mu}$. NH is taken as the true hierarchy.}
\label{posc-a3}
\end{figure}

We perform an exact numerical calculation of the neutrino oscillation probabilities within the framework of 
three-generation mixing and invisible decay of $\nu_3$. The oscillation probabilities are computed in the 
presence of earth matter assuming the PREM density profile \cite{prem}. The oscillation probabilities 
$P_{\mu\mu}$, $P_{e\mu}$, $\bar{P}_{\mu\mu}$ and $\bar{P}_{e\mu}$ as a function of neutrino energy 
for the baseline $L$ = 9700 km, for various values of the decay parameter $\alpha_3$ and $\theta_{23}$
are shown in Fig.~\ref{posc-a3}. The following values have been used to generate these. 
\begin{enumerate}
 \item  ${\delta_{\rm CP}}~=~0^\circ$
 \item  $\theta_{12}~=~34.08^\circ$; $\theta_{23}~=~39^\circ,~45^\circ,~51^\circ$; $\theta_{13}~=~8.5^\circ$
 \item  $\Delta{m^2_{21}}~=~7.6\times10^{-5} (\rm{eV^2})$; $|\Delta{m^2_{32}}|~=~2.4\times10^{-3} (\rm{eV^2})$
 \item  $\alpha_3~=~0,~10^{-4},~10^{-3}~(\rm{eV^2})$
\end{enumerate}
First let us consider the effect of $\alpha_3$ alone for a given $\theta_{23}$. 
The plots for $\alpha_3$ = 0 correspond to the oscillation only case and as the 
value of $\alpha_3$ increases the effect of decay becomes prominent which can be seen from the figure. 
%One can understand the behavior of the 
%oscillation probabilities for different choices of $\alpha_3$ from Fig.~\ref{exp-aLbE-all}. 
%We had seen that 
%for $L=9700$ km, the relevant range of $\alpha_3$ that gives significant decay is the range $\alpha_3 > 10^{-5}$ eV$^2$.
In general the effect of decay is seen to be more for the lower energy neutrinos.
%is again seen in Fig.~\ref{posc-a3}. 
For the decay constant $\alpha_3 = 10^{-4}$~eV$^2$, the effect of
decay increases and the neutrino probabilities show significant depletion 
as compared to the no decay case 
for neutrino energies up to $\sim 15$~GeV. The presence of decay reduces the oscillation amplitude 
near maxima and elevates it near minima.
As $\alpha_3$ increases to $10^{-3}$~eV$^2$, the survival
probability of the neutrino and anti-neutrinos show a difference over the entire energy range considered.
We also note that the effect of decay is mainly to damp out the oscillatory behavior in the 
probabilities. For the large decay case the oscillatory behavior is seen to be largely washed out.  
%probabilities $P_{e\mu}$ and $\bar{P}_{e\mu}$.
From Fig.~\ref{posc-a3} it can be seen that, the relative change in the 
oscillation probability due to decay is more for $\bar{P}_{\mu\mu}$ than $P_{\mu\mu}$ whereas 
the relative change in $P_{e\mu}$ is more compared to that in $\bar{P}_{e\mu}$.
Hence the contribution to the $\alpha_3$ sensitivity $\chi^2$ will be more from anti-neutrino events in the former 
case and neutrino events in the latter case. However since $P_{\mu\mu}$ and $\bar{P}_{\mu\mu}$ are the
dominant channels at ICAL, the major contribution to $\alpha_3$ sensitivity is expected to come 
from anti-neutrino events in the present study.

Now let us look at the effect of $\theta_{23}$ alone for a given $\alpha_3$ value. The effect of
$\theta_{23}$ is also to vary the oscillation amplitude. In general, $P_{\mu\mu}$ and $\bar{P}_{\mu\mu}$
decrease with increase in $\theta_{23}$. However beyond 13~GeV, for $\alpha_3$ = 0 and $10^{-4}$~eV$^2$,
$\theta_{23}=45^\circ$ gives the lowest probability compared to those for $39^\circ$ and $51^\circ$, 
though the relative variation is much less. From the plots in the lower panels of Fig.~\ref{posc-a3}
we see that $P_{e\mu}$ and $\bar{P}_{e\mu}$ increase with $\theta_{23}$, the increase in 
$P_{e\mu}$ is larger than that in $\bar{P}_{e\mu}$. For all values of $\theta_{23}$, $P_{e\mu}$ and 
$\bar{P}_{e\mu}$ decrease. 

Since both $\alpha_{3}$ and $\theta_{23}$ affect the oscillation amplitudes, when combined in the 
following way, similar probabilities can be obtained. The combination of $\theta_{23}$ in the first 
octant + a larger (smaller) value of $\alpha_{3}$ will give a probability similar to that with 
$\theta_{23}$ in second octant + a smaller (larger) value of $\alpha_3$ for $P_{\mu\mu}$ and 
$\bar{P}_{\mu\mu}$ ($P_{e\mu}$ and $\bar{P}_{e\mu}$). Since the event spectrum is dominated by 
$P_{\mu\mu}$ and $\bar{P}_{\mu\mu}$ events, this combined effect will affect the 
sensitivity/discovery potential to/of $\alpha_3$ and the precision measurement on $\theta_{23}$
,which is discussed in Section~\ref{a3-pre}.

%It can be seen that the invisible decay scenario affects the oscillation 
%probabilities more at low neutrino energies. This is pronounced in the case of $E_\nu$ = 0.5 GeV, where 
%$P_{\mu\mu}$ when $\alpha_3=10^{-5}$ eV$^2$ is significantly different from the no oscillation case. 
%As the energy of the neutrino increases, the effects become less distinguishable in the 
%oscillation probability. Hence, not to
%lose any effect, the analysis is done in the energy range 0.5-25 GeV. 
%The higher energy events beyond 11 GeV till 25 GeV are included even though the number of events is lesser
%due to lesser flux. The lower energy events will contribute mainly to the sensitivity to the decay parameter 
%where as higher energy events will enhance the sensitivity to oscillation parameters other than decay 
%lifetime. 
%% This will be illustrated in detail in \ref{chi2}.

\section{Details of numerical simulations}\label{nu-evt-gen}

ICAL will be a 50 kton magnetised iron detector which is optimised for the detection of atmospheric $\nu_\mu$
and $\bar{\nu}_\mu$.Both $\nu_\mu$ ($\bar{\nu}_\mu$) and $\nu_e$ ($\bar{\nu}_e$) fluxes can contribute to the 
$\nu_\mu$ ($\bar{\nu}_\mu$) events observed at ICAL. Hence the number of events detected by ICAL will be : 
\begin{eqnarray} \nonumber
\frac{d^2N}{d E_\mu d\cos\theta_\mu} & = & t \times {n_d}\times
\int{d E_\nu d\cos\theta_\nu d\phi_\nu} \times \\ 
& & \hspace{0.5cm}
\left[P^m_{\mu\mu} \frac{d^3\Phi_\mu}{d E_\nu d\cos\theta_\nu d\phi_\nu}+
P^m_{e\mu} \frac{d^3\Phi_e}{d E_\nu d\cos\theta_\nu d\phi_\nu}
\right] \times
\frac{d\sigma_\mu (E_\nu)}{d E_\mu d\cos\theta_\mu}~,
\label{toteve}
\end{eqnarray}
where $n_d$ is the number of nucleon targets in the detector, $\sigma_\mu$
is the differential neutrino interaction cross section in terms of the
energy and direction of the muon produced, $\Phi_\mu$ and $\Phi_e$ are the
$\nu_\mu$ and $\nu_e$ fluxes and $P^m_{\alpha\beta}$ is the oscillation
probability of $\nu_\alpha\rightarrow\nu_\beta$ in matter and in presence of decay. 
A sample of 1000 years of 
unoscillated neutrino events are generated using NUANCE-3.5 neutrino generator \cite{nuance},
in which the Honda 3D atmospheric neutrino fluxes \cite{honda3d} along with 
neutrino-nucleus cross-sections and a simplified ICAL detector geometry are incorporated.
Each event is oscillated by multiplying with the relevant oscillation probability
including decay and oscillations in Earth matter assuming PREM density profile \cite{prem}. 
The probabilities are obtained by solving the propagation equation in matter in presence of decay.
%by solving a first order differential equation. 
The events are then smeared according to the resolutions and efficiencies obtained from \cite{mupaper1,mthesis}. 
These two steps are done on an event by event basis for the entire 1000 year sample. Both ``data'' and theory 
are generated via this method, ``data'' with the central values of the parameters as described in 
Table~\ref{osc-par-3sig} and theory by varying them in their respective 3$\sigma$ ranges. Afterwards 
the oscillated samples of 1000 years of events, both ``data'' and theory are scaled down to the required 
number of years, 10 for our current analysis. This is done to reduce the effect of Monte-Carlo fluctuations 
on sensitivity studies.

In the current analysis, the efficiencies and resolutions of muons in the central region of the detector 
\cite{mupaper1,mthesis} have been used over the entire detector.
These resolutions and efficiencies have been obtained by the INO collaboration 
via detailed detector simulations using a GEANT4-based simulation toolkit for ICAL.  
The central region of the ICAL detector \cite{mupaper1, mthesis} has the best efficiencies
and resolutions for muons, the few-GeV muons in ICAL have a momentum resolution 
of $\sim$ 10\% and direction resolution of $\sim$ 1$^\circ$ on the average.
Their relative charge identification efficiencies is about $\sim$ 99\%. 
However, ICAL has two more regions namely 
the peripheral \cite{peripheralmu,kthesis} and side regions depending on the magnitude and strength of the 
magnetic field. The peripheral region which has lesser reconstruction efficiencies but only slightly 
worse resolutions compared to the central region, constitutes ~50\% of the detector. Hence, in a realistic 
scenario where the efficiencies and resolutions in different regions are taken appropriately, the results 
obtained with 10 years of running of 50 kton of ICAL will only be obtained by increasing the run time to 
~11.3 years, as mentioned in \cite{hi-mu}.

%\begin{eqnarray} 
%N^{tot}_{\mu^-}(E^{obs}_\mu,\cos\theta^{obs}_\mu) & = &
%N_{\mu^-} \epsilon_{rec} \epsilon_{cid} +
%N_{\mu^+} \epsilon_{rec} (1-\epsilon_{cid})~, \\ \nonumber 
%N^{tot}_{\mu^+}(E^{obs}_\mu,\cos\theta^{obs}_\mu) & = &
%N_{\mu^+} \epsilon_{rec} \epsilon_{cid} + 
%N_{\mu^-} \epsilon_{rec} (1-\epsilon_{cid})~,
%\end{eqnarray}
%where $N^{tot}_{\mu^-}$ ($N^{tot}_{\mu^+}$) is the total number
%of oscillated $\nu_\mu$ ($\overline{\nu}_\mu$) charged current muon neutrino 
%events observed in an $(E^{obs}_\mu,\cos\theta^{obs}_\mu)$ bin;
%$\epsilon_{rec}$ and $\epsilon_{cid}$ are the reconstruction efficiency and 
%relative charge identification efficiency of muons which 
%are a function of the true energy as well as 
%zenith angle direction of the muons. 
Since the charged current $\nu_\mu$ ($\bar{\nu}_\mu$) interactions have 
$\mu^-$ ($\mu^+$) in the final state along with the hadron shower, and since 
ICAL is capable of measuring the energy of the hadron shower, we include in our 
analysis the data on those as well. It was reported in \cite{hres1} from ICAL simulations that hadrons 
in ICAL have energy resolutions of 85\% at 1~GeV and
36\% at 15~GeV and the events are smeared accordingly before including them in 
the final 3D-binned analysis which includes muons binned in observed energy and direction
and hadrons binned in energy. There are 15 bins in $E^{obs}_\mu$ between ($0.5-25$)~GeV, 21 bins in 
$\cos\theta^{obs}_\mu$ between ($-1,+1$)and 4 bins in $E'^{obs}_{had}$ between ($0-15$)~GeV, thus giving 1260 bins. 
More details of the binning scheme and the numerical simulations can be found in Ref.~\cite{hi-mu}.

The true values and the 3$\sigma$ ranges of the oscillation parameters used to generate the probabilities
are given in Table~\ref{osc-par-3sig}. Since 
ICAL is not directly sensitive to $\delta_{CP}$, it is taken as $0^\circ$ in this analysis and kept 
fixed. The 1-2 oscillation parameters $\Delta m^2_{21}$ and 
$\sin^2 \theta_{12}$ are also kept fixed throughout our analysis.
For the remaining parameters two types of analyses are performed,
--- one with fixed parameter 
and the other with marginalisation. In the former all parameters are kept fixed while in the latter, 
the parameters other than the one for which the sensitivity study is done are marginalised in their 
respective 3$\sigma$ ranges shown in Table~\ref{osc-par-3sig}.  

\begin{table}[htp]
\centering 
\begin{tabular}{|c|c|c|}
\hline
 Parameter & True value & Marginalization range \\
\hline
$\theta_{13}$ & 8.5$^\circ$ & [7.80$^\circ$, 9.11$^\circ$] \\
$\sin^{2}\theta_{23}$ & 0.5 & [0.39, 0.64] \\
$\Delta{m^2_{32}}$ & $2.366\times10^{-3}~{\rm eV}^2$ & [2.3, 2.6]$\times10^{-3}~{\rm eV}^2$ (NH) \\
% $\alpha_3$ & 0 eV$^2$ & [0, 2.35$\times10^{-4}$ eV$^2$] \\
$\sin^{2}\theta_{12}$ & 0.304 & Not marginalised \\
$\Delta{m^{2}_{21}}$ & $7.6\times10^{-5}~{\rm eV}^2$ & Not marginalised \\
$\delta_{CP}$ & 0$^\circ$ & Not marginalised \\
\hline    
\end{tabular}
\caption{Oscillation parameters used in this analysis. For fixed parameter studies all parameters are kept at 
their true values. While applying marginalisation, only 
the parameter for which the sensitivity study is being performed is kept fixed and the others are varied in their 
respective 3$\sigma$ ranges.}
\label{osc-par-3sig} 
\end{table}

\noindent
To statistically analyse the data, we define the following $\chi^2$ function
\begin{eqnarray} 
\chi^2 & = & {\stackrel{\hbox{min}}{\displaystyle \xi_l^{\pm},\xi_6}} 
\sum^{N_{E^{obs}_{\mu}}}_{i=1}\sum^{N_{\cos\theta^{obs}_{\mu}}}_{j=1}
\sum^{N_{E'^{obs}_{had}}}_{k=1} 2\left[ \left(T^{+}_{ijk}
- D^{+}_{ijk} \right) - D^{+}_{ijk} \ln \left( 
\frac{T^{+}_{ijk}}{D^{+}_{ijk}} \right) \right] + \nonumber \\
& & 2\left[\left(T^{-}_{ijk} - D^{-}_{ijk}\right) - D^{-}_{ijk}
\ln\left(\frac{T^{-}_{ijk}}{D^{-}_{ijk}}\right)\right] + 
\sum^{5}_{l^{+}=1} \xi^{2}_{l^{+}} + \sum^{5}_{l^{-}=1}
\xi^{2}_{l^{-}} + \xi^{2}_6~.
\label{chisq-11p}
\end{eqnarray}
Here $i,j,k$ sum over muon energy, muon angle and hadron energy bins respectively. The number of 
predicted (theory) events with systematic errors in each bin are given by 
% \begin{eqnarray}
\begin{equation}
 T^{+}_{ijk} = T^{0+}_{ijk}\left(1+\sum^{5}_{l^{+}=1} 
\pi^{l^{+}}_{ijk}\xi_{l^{+}}+\pi_6\xi_6\right)~;
T^{-}_{ijk} = T^{0-}_{ijk}\left(1+\sum^{5}_{l^{-}=1}
\pi^{l^{-}}_{ijk}\xi_{l^{-}}-\pi_6\xi_6\right)~.
\label{td-pi6xi6}
\end{equation}
% \end{eqnarray}
The number of theory events without systematic errors in a bin is given by $T^{0\pm}_{ijk}$ and the observed 
events (``data'') per bin are given by $D^{\pm}_{ijk}$. It should be noted that
both $D^{\pm}_{ijk}$ and $T^{0\pm}_{ijk}$ are obtained from the scaled NUANCE neutrino
events as mentioned earlier. The following values are taken for the systematic uncertainties 
\cite{kameda,maltoni}: $\pi_1=20$\% flux normalisation error, $\pi_2=10$\% cross section error,
$\pi_3=5$\% tilt error, $\pi_4=5$\% zenith angle error, $\pi_5=5$\% overall systematics
and $\pi_6=2.5$\% on $\Phi_{\nu_\mu}/\Phi_{\bar{\nu}_\mu}$ ratio. These are included in the analysis via 
pull method. The ``tilt'' error is incorporated as follows. The event spectrum with the predicted values 
of atmospheric neutrino fluxes is calculated and then shifted according to the relation :
\begin{equation}
\Phi_\delta(E) = \Phi_0(E)(\frac{E}{E_0})^\delta \simeq \Phi_0(E)(1+\delta\ln\frac{E}{E_0}),
\end{equation}
where $E_0$ is 2~GeV, and $\delta$ is the 1$\sigma$ systematic tilt error (5\%). Flux error is 
included as the difference $\Phi_\delta(E)-\Phi_0(E)$.

A prior of 8\% at 1$\sigma$ is added to $\sin^22\theta_{13}$. This is the only prior in this 
calculation. No prior is imposed at all on the quantities whose sensitivities are to be 
studied, i.e on $\alpha_3, \theta_{23}$ and $|\Delta{m^2_{32}}|$.
The contribution from prior to the $\chi^2$ is :
\begin{equation}
\chi_{\rm prior}^2 = \left(\frac{\sin^22\theta_{13}-
\sin^22\theta_{13}^{\rm true}}
{\sigma(\sin^22\theta_{13})}\right)^2~,
\label{chi2-prior}
\end{equation}
where, $\sigma(\sin^22\theta_{13})=0.08\times\sin^22\theta^{\rm true}_{13}$.
Hence, the final $\chi^2$ for ICAL 
will be :
\begin{equation}
\chi_{\rm ICAL}^{2} = \chi^2+\chi_{\rm prior}^2~,
\label{chi2-ical}
\end{equation}
where $\chi^2$ is given by Eq.~(\ref{chisq-11p}).

%%%%%%%%%%%%%%%%%%%%%%%%%%%%%%%%%%%%%%%%%%%%%%%%%%%%%%%%%%%%%%%%%%%%%%%%%%%%%%%%%%%%%%%%%%%%%%%%%%%%%%%%%%%%%%
\section{Sensitivity of ICAL to $\alpha_3$}\label{a3-results}
   The results of the sensitivity studies of ICAL to $\alpha_3$ are presented in this section. 
   We first show how the number of oscillated events change with decay as a function of 
   zenith angle and energy. Then we proceed further to discuss the sensitivity as well as the 
   discovery potential of ICAL to neutrino decay and the bound on $\alpha_3$ from our analysis. 
   
\begin{figure}[htp]
\includegraphics[width=0.42\textwidth,height=0.4\textwidth]{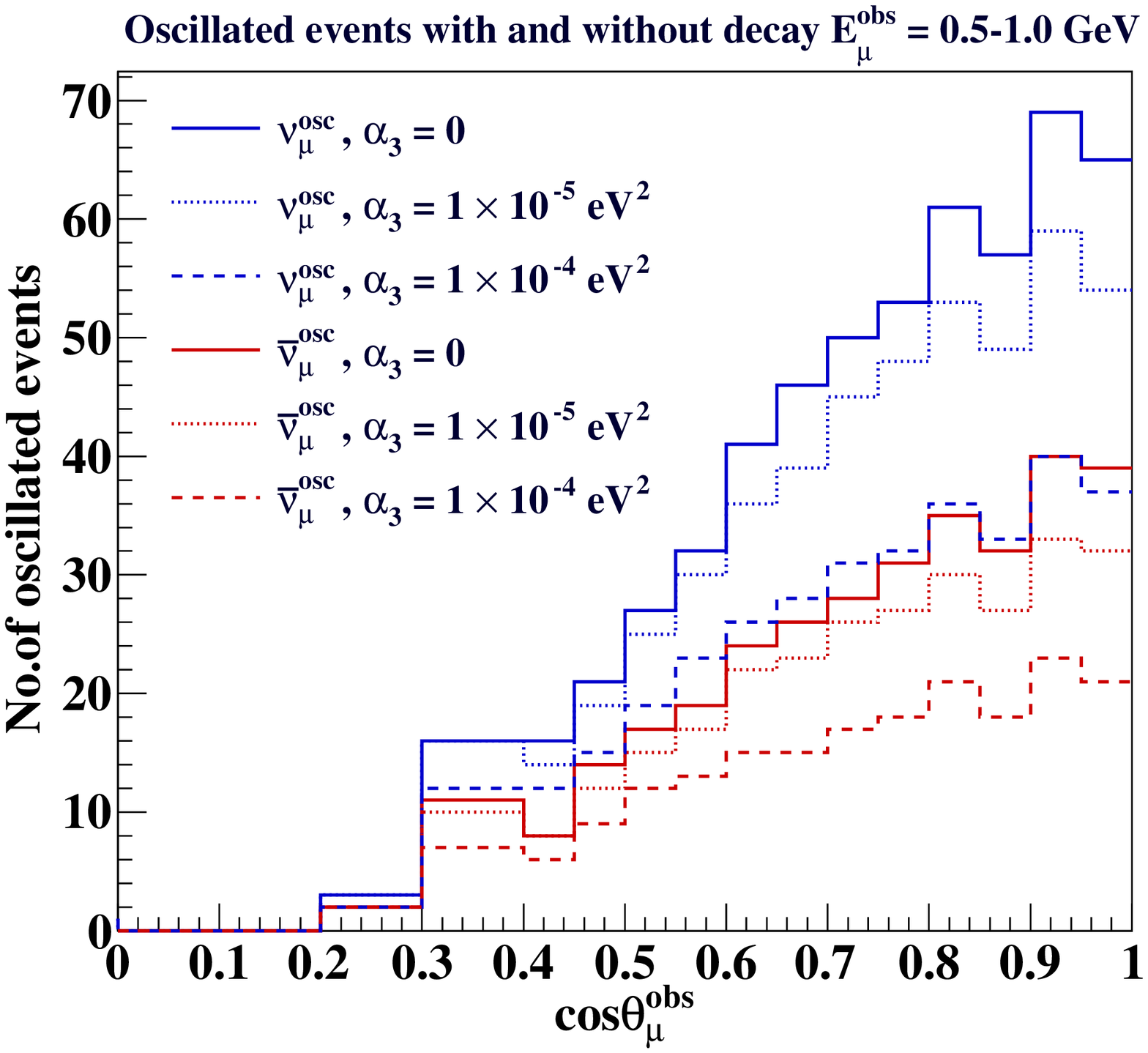}
\includegraphics[width=0.42\textwidth,height=0.4\textwidth]{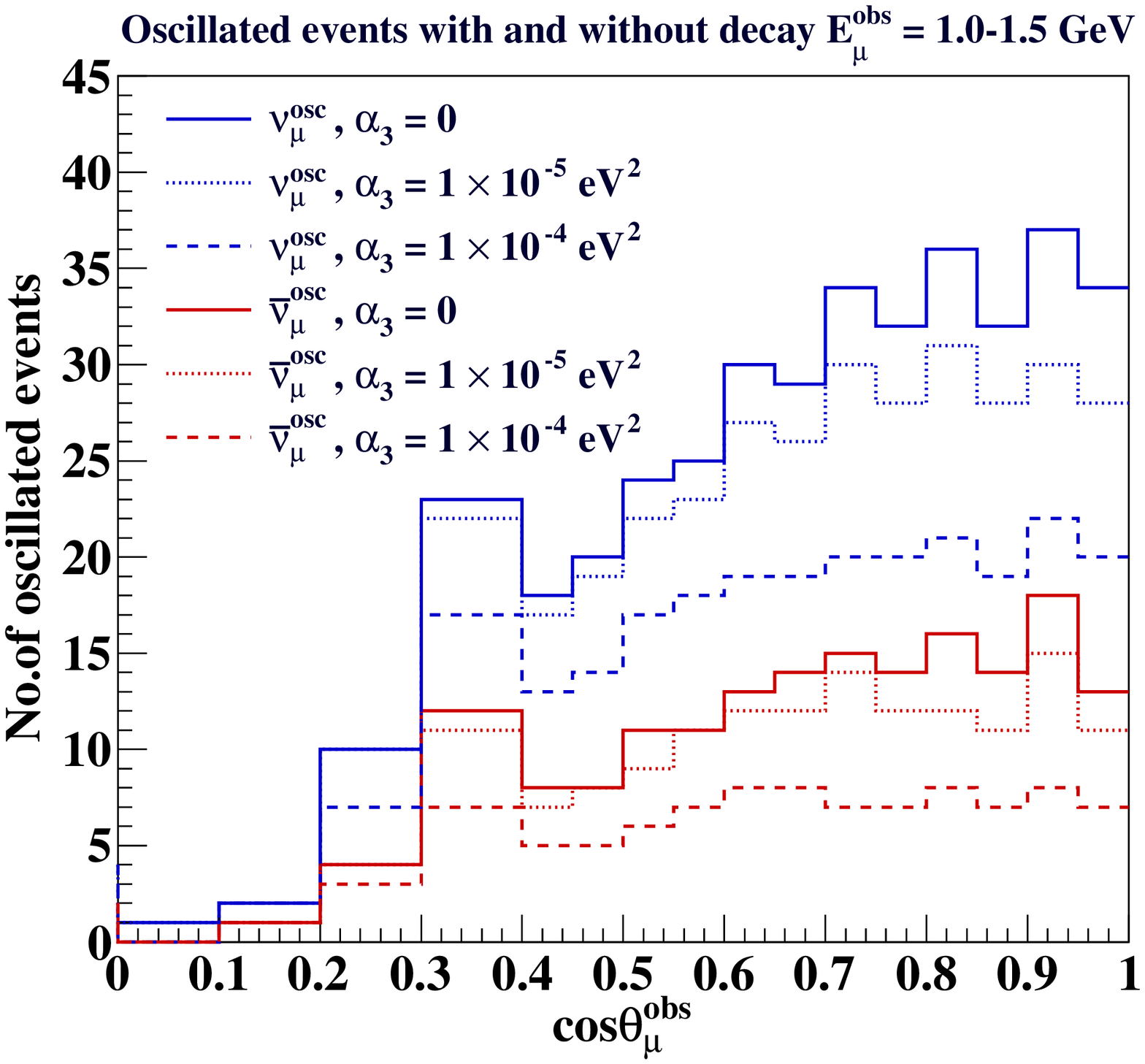}
\includegraphics[width=0.42\textwidth,height=0.4\textwidth]{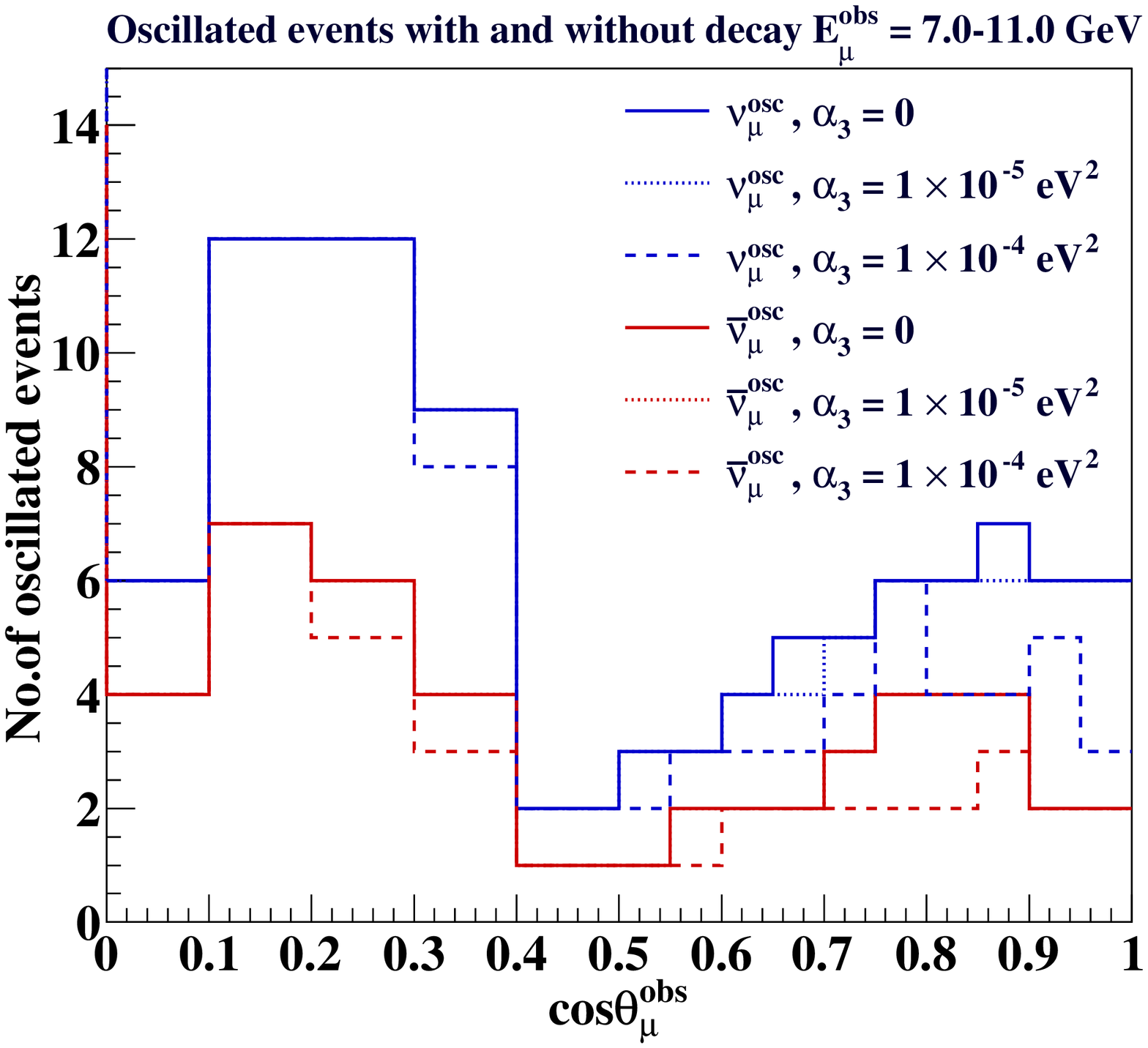}
\includegraphics[width=0.42\textwidth,height=0.4\textwidth]{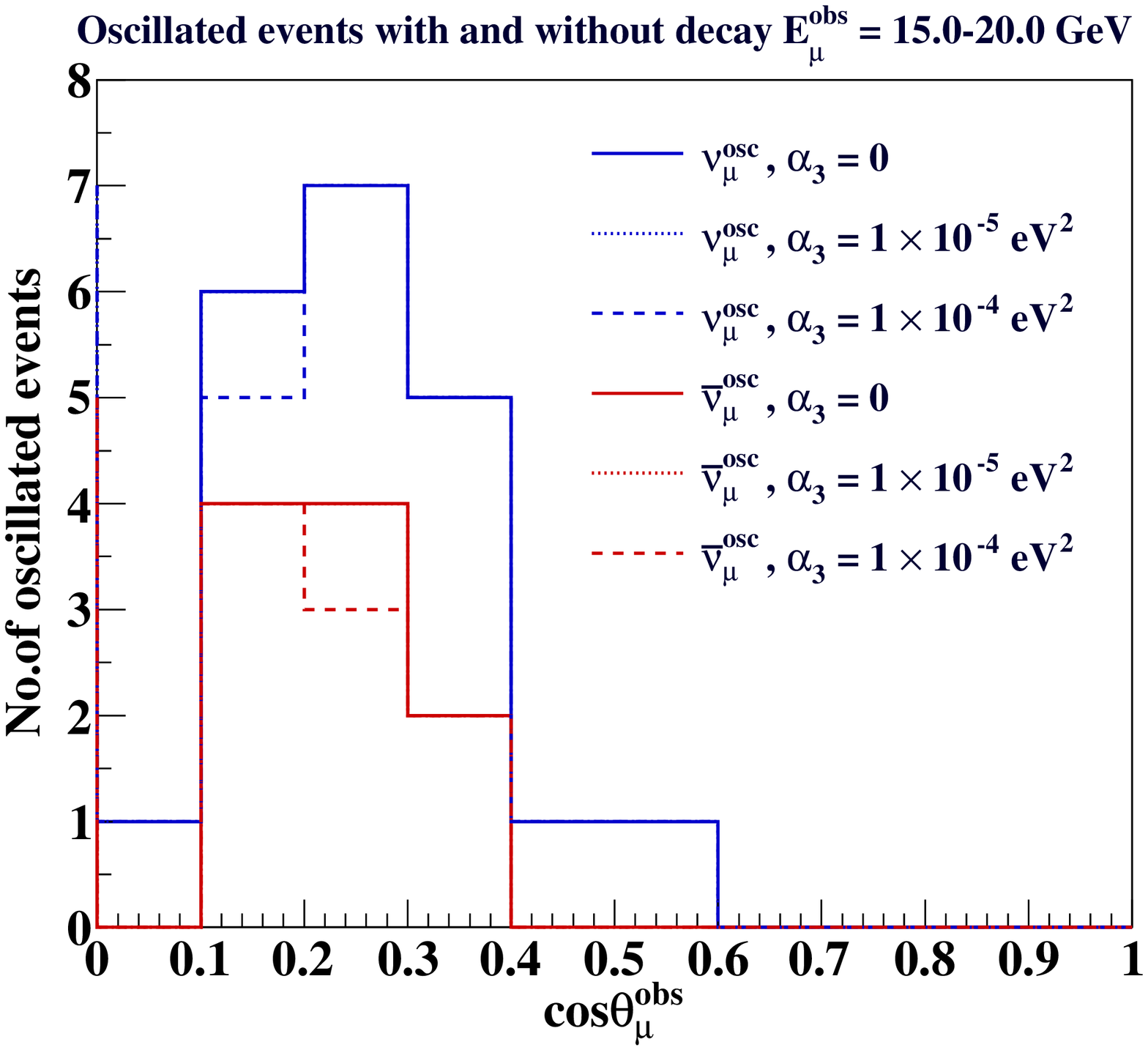}
\caption{Oscillated $\nu_\mu$ and $\bar{\nu}_\mu$ events for each $E^{obs}_\mu$ bin as a function of 
$\cos\theta^{obs}_\mu$ for $\alpha_3=0,1\times10^{-5}$ and $1\times10^{-4}$~eV$^2$ . The other parameters are set 
to their central values as in Table~\ref{osc-par-3sig}. It should be noted that the y-axes are not the same. Only
up-coming events (oscillated) are shown here.}
\label{nevt-osc-a3}
\end{figure}

\paragraph{Effect of decay on the number of oscillated events:}\label{a3-effect}
In Fig.~\ref{nevt-osc-a3}, we show  the zenith angle distribution of the 
$\nu_\mu$ and $\bar\nu_\mu$ 
events for different values of the decay constant $\alpha_3$. 
The four panels are for four 
different energy bins. 
The convention used in these plots is such that $\cos\theta^{obs}_\mu=[0,1]$ indicates the up-coming neutrinos. 
It can be seen from the figure that both $\nu_\mu$ and $\bar\nu_\mu$ events deplete with an increase in the value of 
 $\alpha_3$. We also note that the effect of decay is more prominent in the lower energy bins. With increase in 
 energy, there is no significant effect of decay on the number of events if the decay parameter is less than
 $10^{-4}$~eV$^2$ 
as can be seen from the lower panels. 

\begin{figure}[htp]
 \includegraphics[width=0.42\textwidth,height=0.40\textwidth]{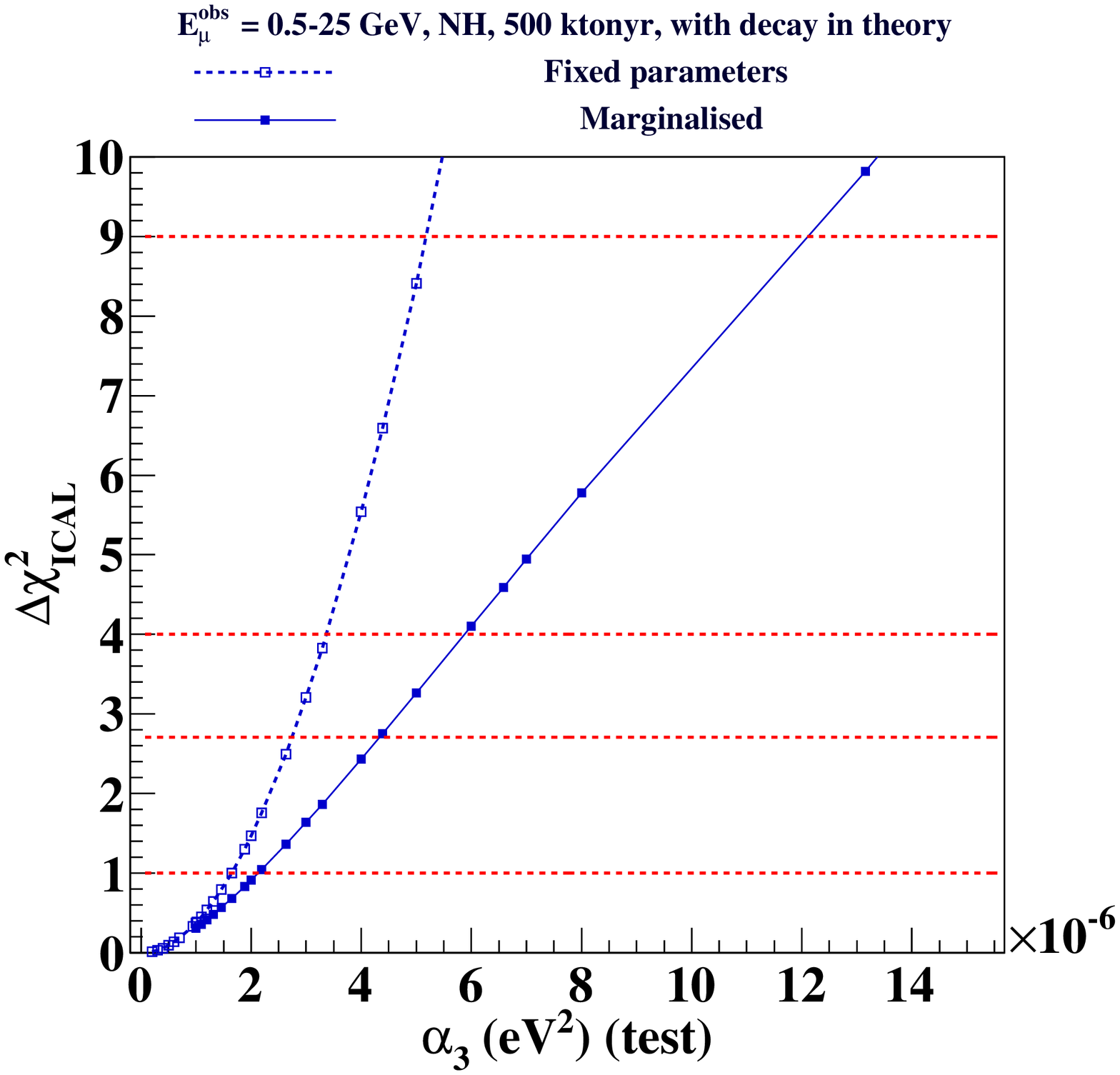} 
 \includegraphics[width=0.42\textwidth,height=0.40\textwidth]{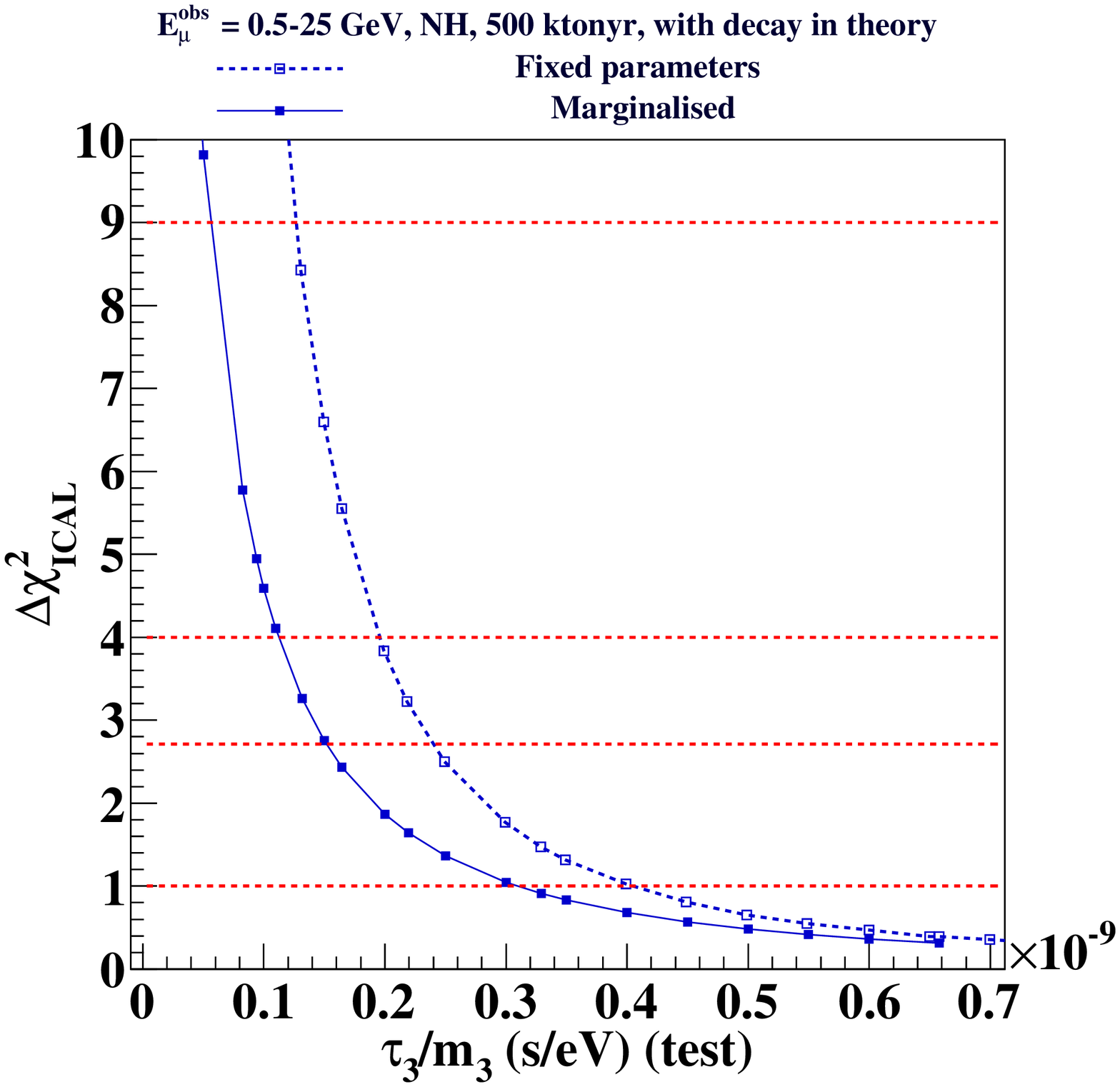}
 \caption{Expected sensitivity of ICAL to neutrino decay. The expected $\chi^2$ is shown as a function of $\alpha_3$
 (test) eV$^2$ (left panel) and $\tau_3/m_3$(test)~(s/eV) (right panel) 
 with 500 kton-yr exposure of ICAL. }
 \label{a3-sensitivity}
\end{figure}

\clearpage
\paragraph{Sensitivity to the decay parameter $\alpha_3$:}\label{alpha3-sense}
In this section, first the study of the  sensitivity of ICAL to $\alpha_3$ is presented with 500 kton-yr exposure
of the detector taking normal hierarchy (NH) as the true hierarchy. To that end, we simulate the prospective 
``data'' for no decay and fit it with a theory of oscillation plus decay. The corresponding $\chi^2$ is shown as a function
of $\alpha_3$(test) in the left panel of Fig.~\ref{a3-sensitivity}.

The blue dashed curve is obtained for a fixed parameter fit while the blue solid one
corresponds to the sensitivity when the $\chi^2$ is marginalised over all oscillation parameters as described in 
Section~\ref{nu-evt-gen}. A comparison of the solid and dashed curves gives us an idea of the impact of 
marginalisation over the oscillation parameters on the sensitivity of the experiment to decay. From 
Fig.~\ref{a3-sensitivity} it can be seen that with marginalisation of the oscillation parameters, the sensitivity 
decreases as expected. The right panel shows the sensitivity to decay in terms of $\tau_3/m_3$ in~s/eV. The 
expected sensitivity of ICAL to $\alpha_3$ are shown in Table~\ref{a3-sense-wh-11p-fix-marg}. The corresponding 
values of $\tau_3/m_3$ in units of~s/eV are also given. Note that by sensitivity limit we mean the 
value of $\alpha_3$ ($\tau_3/m_3$) upto which ICAL can rule out neutrino decay.
   \begin{table}[htp]
  \centering
   \begin{tabular}{|c|c|c|c|}
   \hline
    Analysis type & $\chi^2$ & $\alpha_3$~(eV$^2$) & $\tau_3/m_3$~(s/eV) \\
   \hline
                     & 1	& 1.65$\times10^{-6}$ & 3.99$\times10^{-10}$ \\
		     & 2.71	& 2.73$\times10^{-6}$ & 2.39$\times10^{-10}$ \\
    Fixed parameters & 4	& 3.37$\times10^{-6}$ & 1.96$\times10^{-10}$ \\
		     & 9	& 5.19$\times10^{-6}$ & 1.28$\times10^{-10}$ \\   
    \hline
		     & 1	& 2.13$\times10^{-6}$ & 3.03$\times10^{-10}$ \\
		     & 2.71	& 4.36$\times10^{-6}$ & 1.51$\times10^{-10}$ \\
    Marginalised     & 4	& 5.89$\times10^{-6}$ & 1.12$\times10^{-10}$ \\
		     & 9	& 1.21$\times10^{-5}$ & 5.66$\times10^{-11}$ \\   
    \hline
    \end{tabular}
 \caption{Sensitivity to $\alpha_3$~(eV$^2$) and $\tau_3/m_3$~(s/eV) with 500 kton year exposure of ICAL 
 assuming NH as the true hierarchy. }
 \label{a3-sense-wh-11p-fix-marg}
  \end{table}
\clearpage

The lower bound on $\tau_3/m_3$ for the invisible decay scenario from MINOS data was shown to be 
$\tau_3/m_3>2.8\times10^{-12}$~(s/eV) at 90\% C.L. This corresponds to an upper limit 
$\alpha_3<2.35\times10^{-4}$~eV$^2$. Table \ref{a3-sense-wh-11p-fix-marg} shows that ICAL is
expected to tighten these bounds by two orders of magnitude with just charged current $\nu_\mu$ and 
$\bar{\nu}_\mu$ events. At 90\% C.L, ICAL with marginalisation is expected to give a lower bound of 
$\tau_3/m_3>1.51\times10^{-10}$~(s/eV) which corresponds to $\alpha_3<4.36\times10^{-6}$~eV$^2$. 

\begin{figure}[htp]
 \includegraphics[width=0.42\textwidth,height=0.40\textwidth]{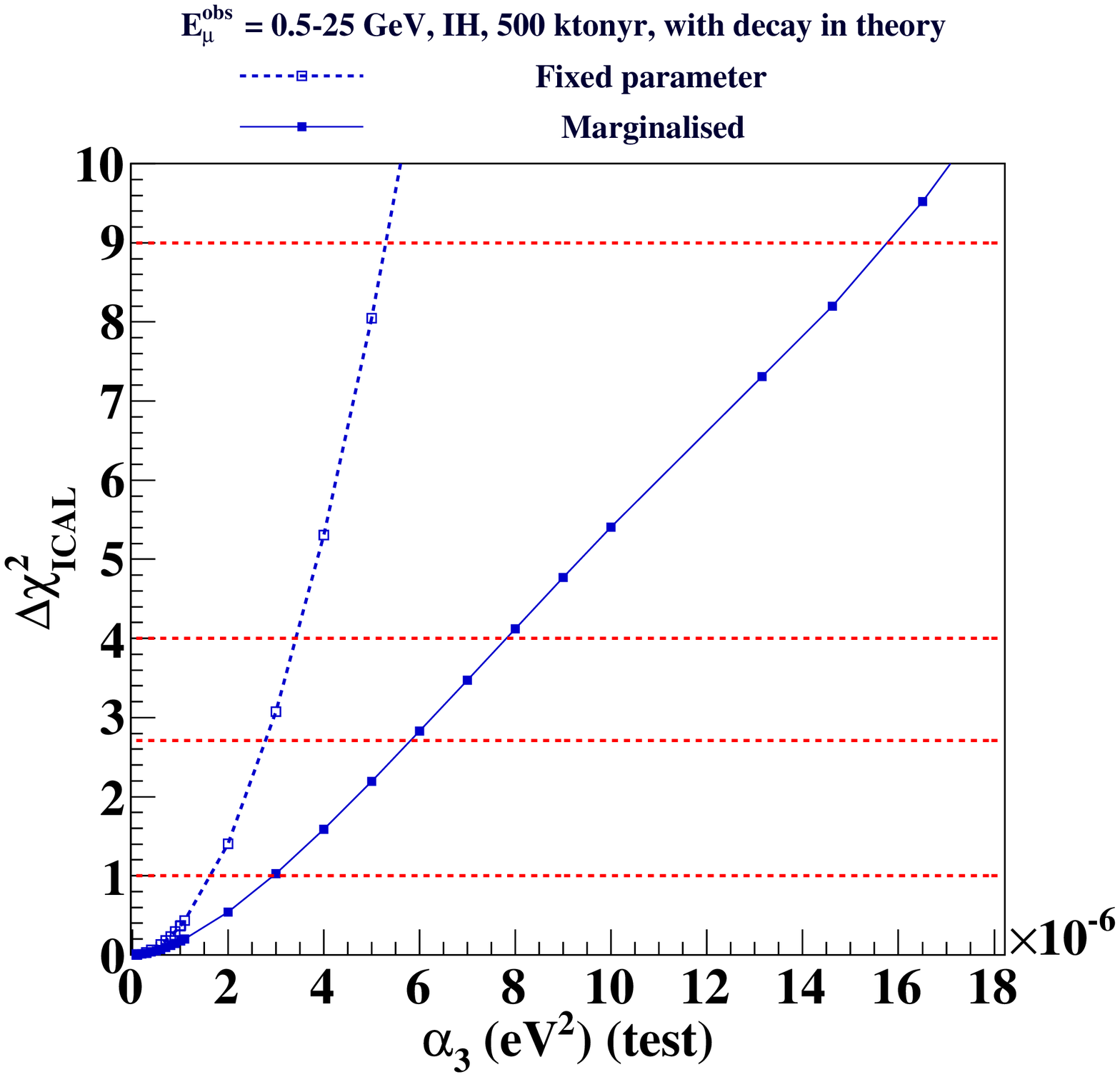} 
 \includegraphics[width=0.42\textwidth,height=0.40\textwidth]{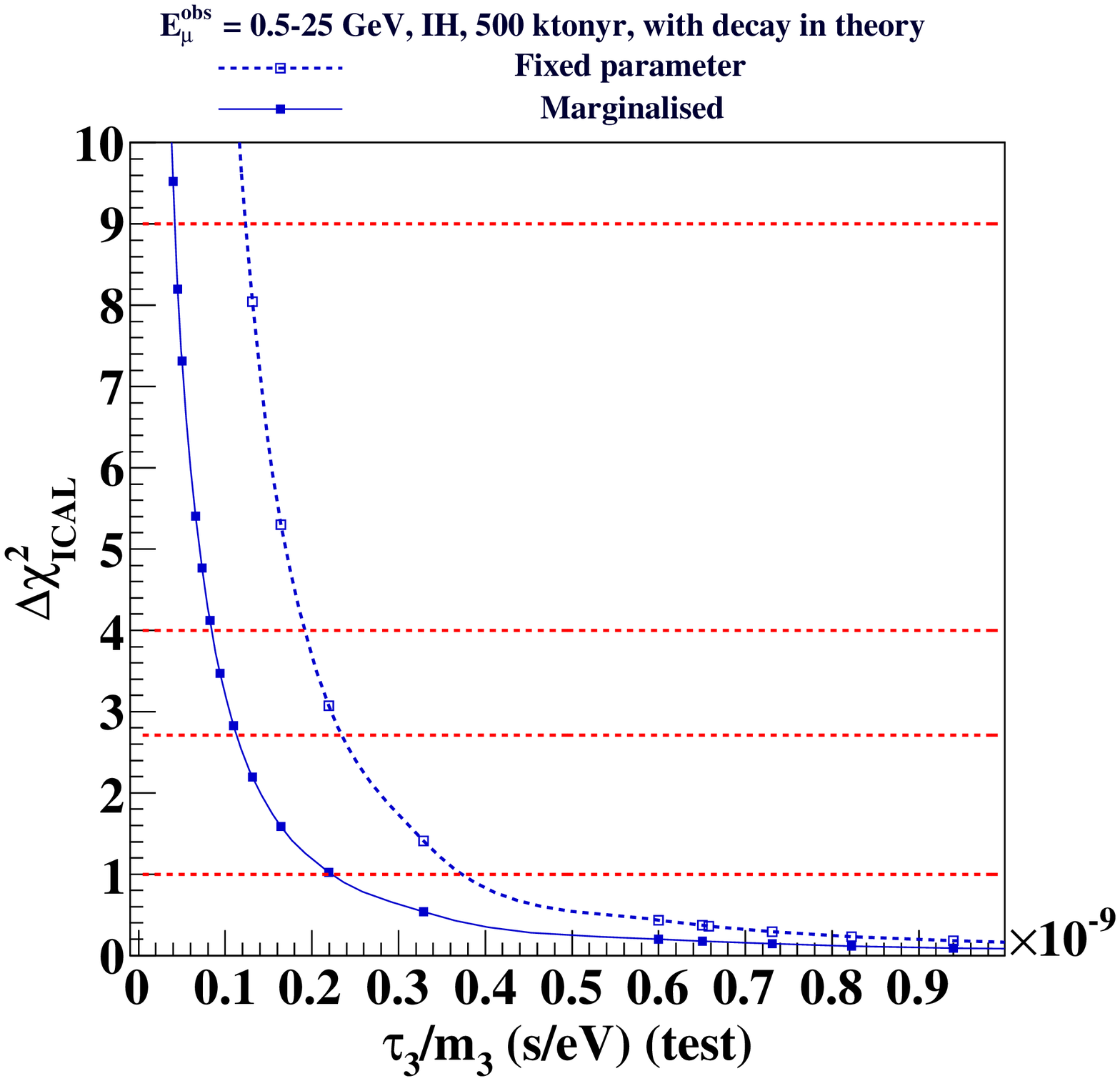}
 \caption{Bounds on the allowed values of (left) $\alpha_3$~eV$^2$ (right) $\tau_3/m_3$~(s/eV) with 500 kton
 year exposure of ICAL with IH as true hierarchy. The comparison of results for fixed parameter and 
 marginalised cases is shown.}
 \label{a3-ih}
\end{figure}

The expected sensitivity with fixed parameters as well as marginalisation for true IH are shown in Fig.~\ref{a3-ih}.  
At 90\% C.L, the upper bound on $\alpha_3$ are $\alpha_3<2.78\times10^{-6}$~eV$^2$ with fixed parameters and 
$\alpha_3<5.82\times10^{-6}$~eV$^2$ with marginalisation. These are only slightly worse than the sensitivities obtained 
with true NH. In terms of $\tau_3/m_3$, these limits translate as the lower limits $\tau_3/m_3>2.42\times10^{-10}$~s/eV 
and $\tau_3/m_3>1.14\times10^{-10}$~s/eV for the fixed parameter and marginalised cases, respectively.
The expected sensitivity to $\alpha_3$ at different C.L. with true IH is summarised in 
Table~\ref{a3-sense-wh-11p-fix-marg-ih}. 

  \begin{table}[htp]
  \centering
   \begin{tabular}{|c|c|c|c|}
   \hline
    Analysis type & $\chi^2$ & $\alpha_3$ (eV$^2$) & $\tau_3/m_3$ (s/eV) \\
   \hline
                     & 1	& 1.65$\times10^{-6}$ & 4.35$\times10^{-10}$ \\
		     & 2.71	& 2.78$\times10^{-6}$ & 2.42$\times10^{-10}$ \\
    Fixed parameters & 4	& 3.43$\times10^{-6}$ & 1.97$\times10^{-10}$ \\
		     & 9	& 5.31$\times10^{-6}$ & 1.25$\times10^{-10}$ \\   
    \hline
		     & 1	& 2.97$\times10^{-6}$ & 2.21$\times10^{-10}$ \\
		     & 2.71	& 5.82$\times10^{-6}$ & 1.14$\times10^{-10}$ \\
    Marginalised     & 4	& 7.82$\times10^{-6}$ & 8.44$\times10^{-11}$ \\
		     & 9	& 1.58$\times10^{-5}$ & 4.21$\times10^{-11}$ \\   
    \hline
    \end{tabular}
 \caption{Sensitivity to $\alpha_3$~(eV$^2$) and $\tau_3/m_3$~(s/eV) with 500 kton year exposure of ICAL 
 assuming IH as the true hierarchy. }
 \label{a3-sense-wh-11p-fix-marg-ih}
  \end{table}

  \begin{table}[htp]
   \begin{tabular}{|c|c|}
    \hline
    
   \end{tabular}

  \end{table}

\begin{figure}[htp]
  \centering
   \includegraphics[scale=0.5]{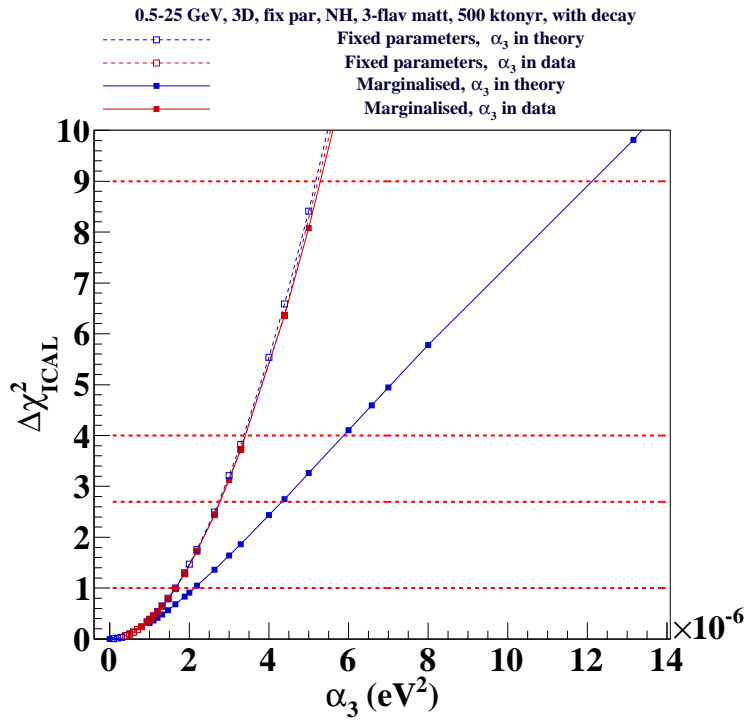}
   \caption{``Discovery potential'' of $\alpha_3$ by ICAL with 500 kton 
   year exposure assuming NH as the true hierarchy, from fixed parameter and marginalised analyses.}
   \label{a3-disc-pot}
   \end{figure}

%\subsection{Discovery potential of $\alpha_3$}
  The analysis discussed above gives us the sensitivity to $\alpha_3$ when we fit 
  a ``data'' with no decay with a theory which has decay. On the other hand, if neutrinos indeed decay into sterile 
  components, and if the decay rate is large enough to be observed in ICAL, we will be able to discover neutrino decay
  at this experiment. Therefore, we next estimate how much the decay rate needs to be in order for ICAL to make 
  this discovery. For this analysis, we simulate the ``data'' with different  values of $\alpha_3$ and fit it with a 
  theory with no decay. The analysis was done for 500 kton-yr exposure of ICAL for fixed parameters  as well as with 
  marginalisation of the undisplayed parameters over their respective 3$\sigma$ ranges.  The results are shown in 
  Fig.~\ref{a3-disc-pot} by the red-dashed curve for the fixed parameter case and the red-solid line for the marginalized 
case. However, we find that for the discovery potential, the marginalization 
has no effect and gives the same result as the fixed parameter case. 
We find that ICAL will be able to discover neutrino decay at the 90\% C.L. if $\alpha_3 > 2.5\times 10^{-6}$~eV$^2$.
We also plot the sensitivity curves, blue dashed (solid) lines for the fixed parameter (marginalized) case, 
in this figure for a comparison between the `sensitivity'' and ``discovery" potential of $\alpha_3$.
We can see that the ``sensitivity" and ``discovery" limits of ICAL are very similar for fixed parameter analysis.
However for the marginalised case the ``discovery potential'' is significantly higher than the ``sensitivity'' 
limit and is same as the fixed parameter case. 
  
\begin{figure}[htp]
\includegraphics[width=0.48\textwidth,height=0.45\textwidth]{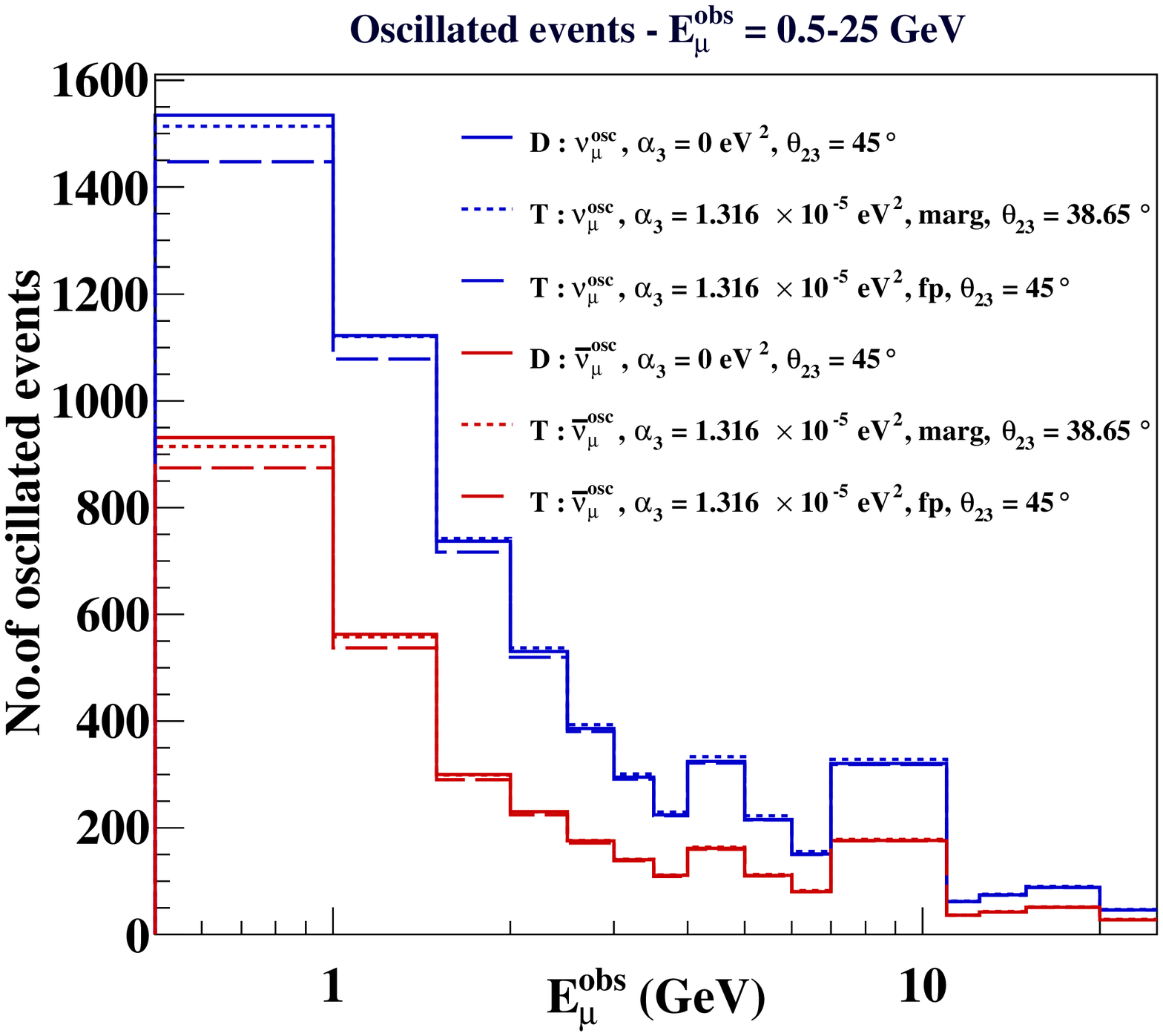}
\includegraphics[width=0.48\textwidth,height=0.45\textwidth]{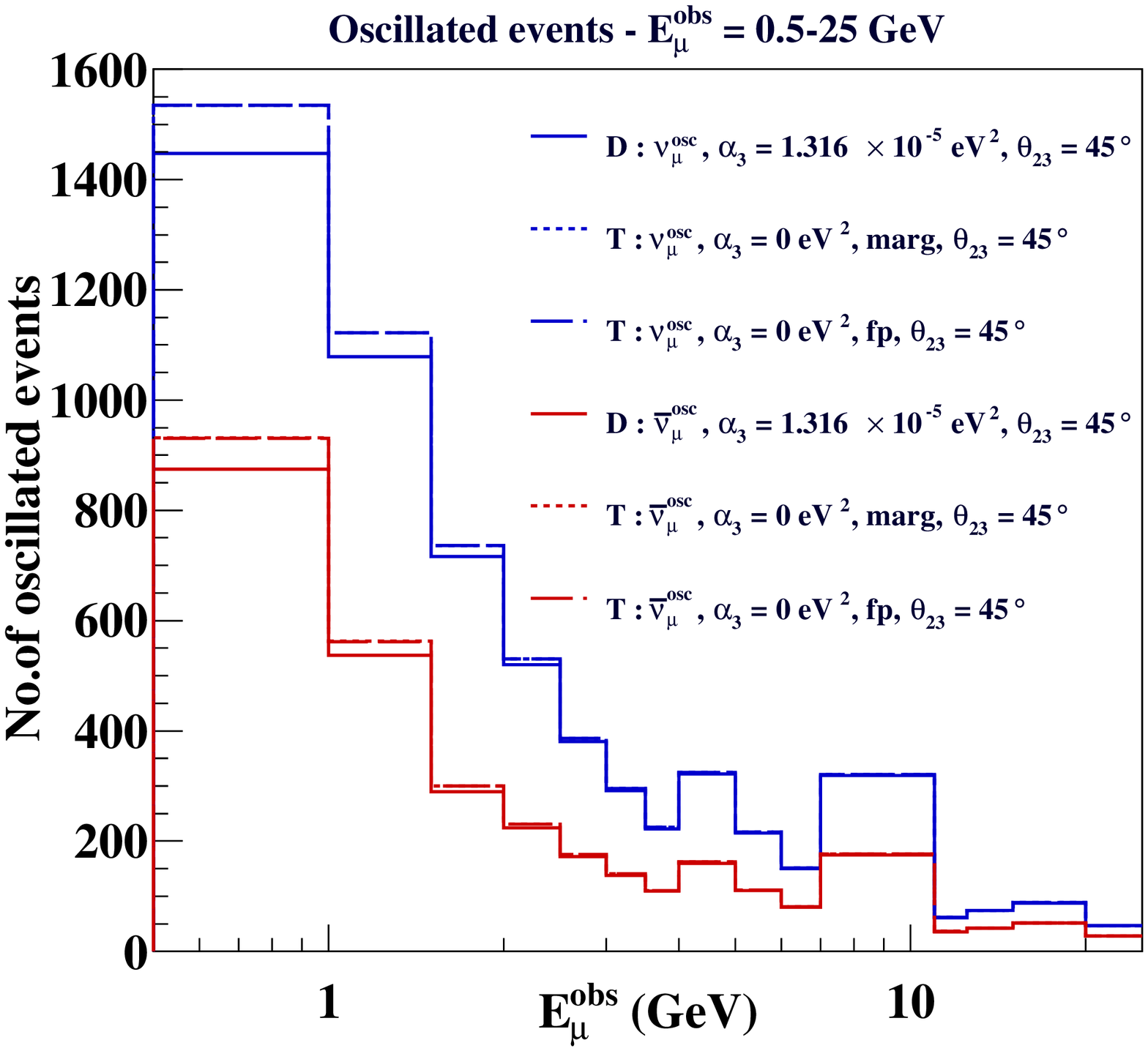}
 \caption{Number of oscillated events per $E^{obs}_\mu$ bin from 0.5--25~GeV (left) with $\alpha_3=0$~eV$^2$
 in ``data'' and $\alpha_3=1.316\times10^{-5}$~eV$^2$ in theory; (right) with $\alpha_3=1.316\times10^{-5}$~eV$^2$
 in ``data'' and $\alpha_3=0$~eV$^2$ in theory for the marginalised case. ``D'' represents data and ``T'' represents 
 theory events. The blue histograms are for $\nu_\mu$ and the red ones are for $\bar{\nu}_\mu$ 
events. The theory events are generated with marginalisation of parameters except $\alpha_3$ in their respective 
3$\sigma$ ranges.}
 \label{evt-mum-mup-a3}
 \end{figure}

The reason why the expected ``sensitivity" limit worsens due to marginalisation while the 
expected ``discovery" limit does not can be understood as follows. 
For the ``sensitivity" analysis we generate the 
data for no decay and $\theta_{23}$ maximal and fit it with a theory where 
$\alpha_3 \neq 0$. Since the effect of 
decay is to reduce the number of events and suppress the event spectrum 
for fixed parameter there will be a difference between the 
data and the theory giving a higher $\chi^2$. 
For the marginalized case, this can be compensated to some 
extent by suitably changing the value of $\theta_{23}$ from maximal and thereby reducing $\sin^22\theta_{23}$, the leading 
term that controls the amplitude of oscillations in the case of muon neutrino survival probability. 
This can be seen in Fig.~\ref{evt-mum-mup-a3}. In this figure the solid line denotes the ``data" generated with
$\alpha_3=0$ i.e no decay and $\theta_{23}= 45^\circ$ while the dashed (dotted) lines show the theory events for 
a non-zero $\alpha_3$ and $\theta_{23}= 45^\circ (38.65^\circ)$. We can see that the lower value of $\theta_{23}$ 
compensates for the depletion due to decay and can give a lower $\chi^2$. As a result the expected sensitivity drops
when the sensitivity $\chi^2$ is marginalised over $\theta_{23}$. 
%%%%%%%%%%%%%%%%%%%%%%%%%%%%%%%%%%%%%%%%%%%%%%%%%%%%%%%%%%%%%%%%%%%%%%%%  
%%%%%%%%%%%%%%%%%%%%%%%%%%%%%%%%%%%%%%%%%%%%%%%%%%%%%%%%%%%%%%%%%%%%%%%

On the other hand, for the expected ``discovery" limit case we generate the 
data for non-zero $\alpha_3$ and maximal mixing and fit it with a theory with no decay. 
In this case, the data has events lower than the theory 
due to decay. This can be seen from the  second panel of 
Fig.~\ref{evt-mum-mup-a3} where the blue (red) solid line 
denotes the data events for muon neutrinos (anti-neutrinos). 
However, unlike the ``sensitivity" case, here one cannot change 
$\theta_{23}$ to reduce the event spectrum any further to compensate for the difference
between data and theory since maximal mixing already corresponds to maximal suppression of 
the muon neutrino survival probability, the leading oscillation channel for atmospheric neutrinos. 
As a result, the fit continues to keep $\theta_{23}$ at its maximal value 
and marginalisation fails to lower the $\chi^2$ any further. 
This can also be seen from Fig.~\ref{evt-mum-mup-a3} where the dotted line shows the theory 
events obtained after marginalizationand this is higher than the data events and same as the fixed
parameter case. 

\section{Precision measurement of $\sin^2\theta_{23}$ and $|\Delta{m^2_{32}}|$}\label{a3-pre}

We next look at the impact of neutrino decay on the precision measurement of 
the mixing angle $\theta_{23}$ and the mass squared difference $|\Delta{m^2_{32}}|$ at ICAL. 
A comparison of the precision measurement in the presence and absence of decay is presented. 
In the no decay case both ``data'' and theory are generated without the decay parameter and in the 
case with decay both ``data'' and theory are generated with non-zero values of $\alpha_3$.
For all results presented in this section, the value $\alpha_3~=~1\times10^{-5}$~eV$^2$ is used to generate the ``data''. 
In the fixed parameter analysis this is kept fixed in theory and for the marginalised case, the range over which $\alpha_3$ 
is marginalised is taken to be $\alpha_3~=~[0,2.35\times10^{-4}]$~eV$^2$ which corresponds to the 90\% CL bound given by the
MINOS analysis. The other parameters are kept fixed at their true values as shown in Table~\ref{osc-par-3sig} for the fixed
parameter analyses and varied in the 3$\sigma$ ranges as shown in the same table for 
the marginalized case. The 1$\sigma$ precision on a parameter 
   $\lambda$ is defined as :
  \begin{equation}
  p(\lambda) = \frac{\lambda_{\hbox{max-}2\sigma}-
  \lambda_{\hbox{min-}2\sigma}}{4\lambda_{true}}~,
  \label{sig1-pre}
  \end{equation}
  where $\lambda_{\hbox{max-}2\sigma}$ and $\lambda_{\hbox{min-}2\sigma}$ are the maximum and minimum allowed 
  values of $\lambda$ at 2$\sigma$ and $\lambda_{true}$ is the true choice. 

\subsection{Precision on $\sin^2\theta_{23}$ in the presence of decay}\label{pre-tt23-wd}
\begin{figure}[htp]
 \includegraphics[width=0.42\textwidth,height=0.40\textwidth]{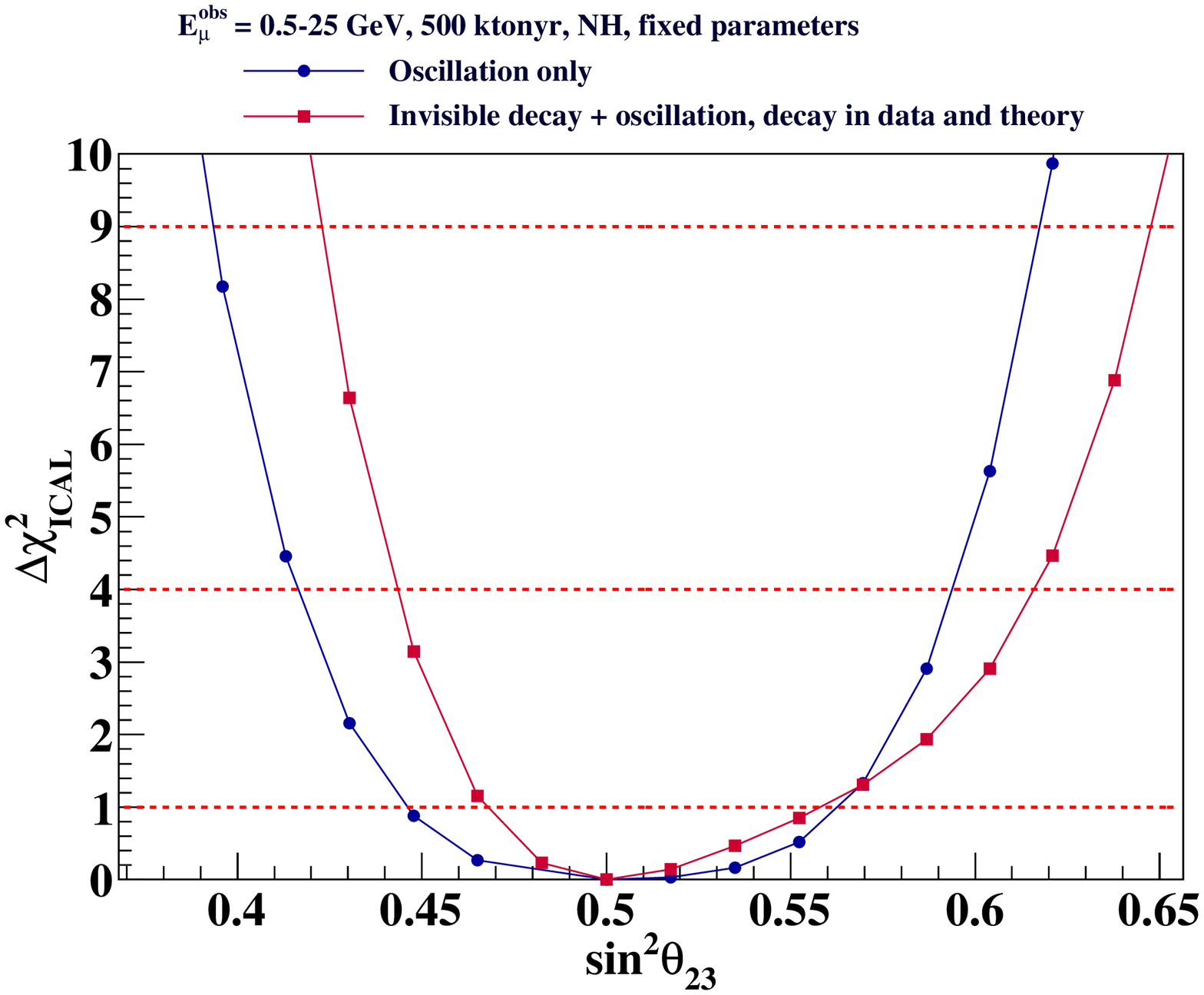}
 \includegraphics[width=0.42\textwidth,height=0.40\textwidth]{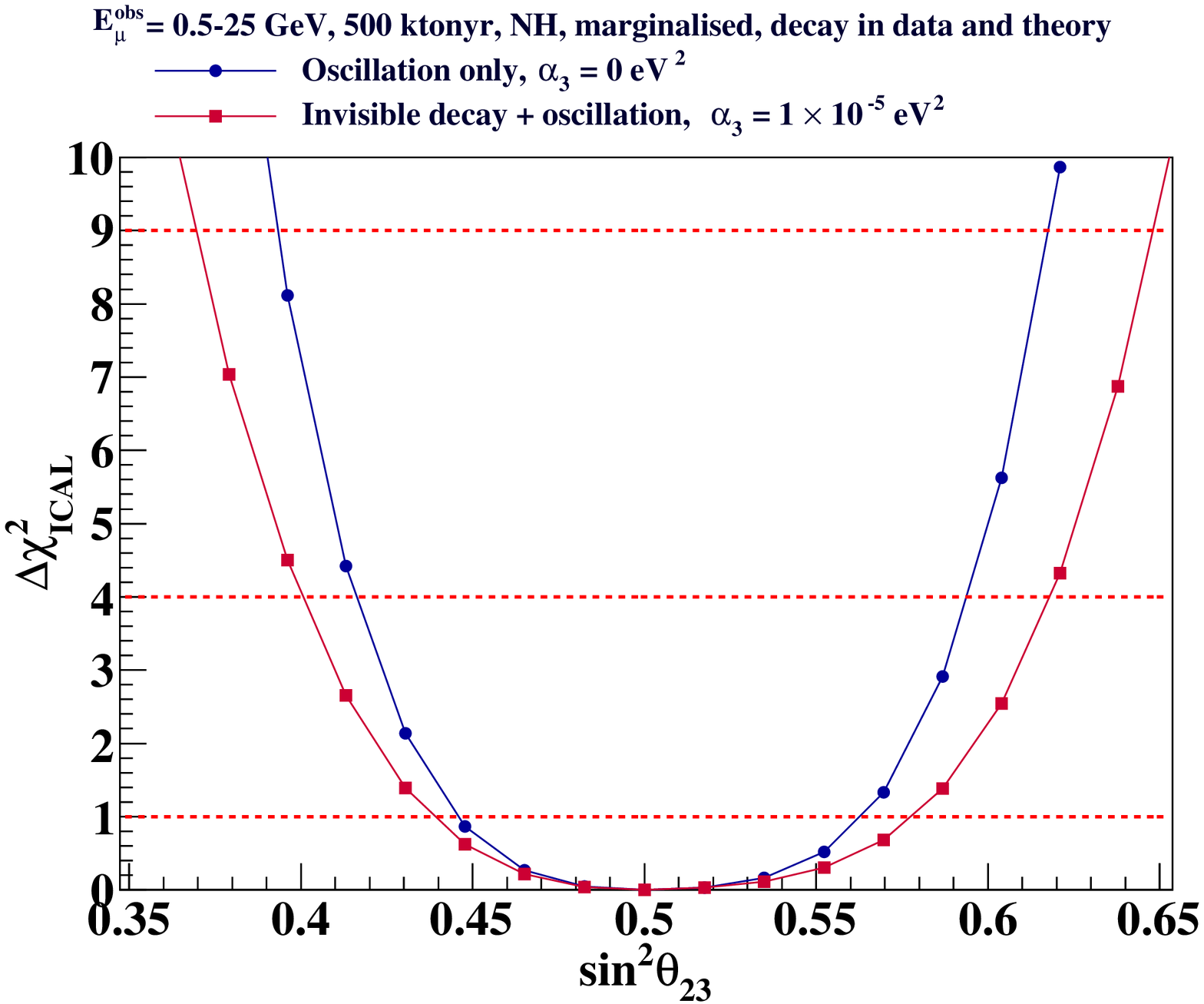}
 \caption{Precision on $\sin^2\theta_{23}$ in the presence and absence of invisible decay for (left)
 fixed parameter case (right) marginalised case. The value of decay parameter $\alpha_3$ in ``data'' is taken to be 
 $1\times10^{-5}$~eV$^2$.}
 \label{tt23-pre-wd}
\end{figure}

:
The sensitivity to $\sin^2\theta_{23}$ in the presence and absence of $\nu_3$ decay is shown 
in Fig.~\ref{tt23-pre-wd}. The left panel shows the fixed parameter results whereas the right panel 
shows the results for the marginalised case. For the fixed parameter case, in the absence of decay, 
the 1$\sigma$ precision on $\sin^2\theta_{23}$ is
$\sim$ 8.9\%. In presence of decay the 1$\sigma$ precision is $\sim$8.6\% which is 
similar to the no decay case. However, it is important to note that even though the percentage precision is same, 
the allowed parameter space is shifted to the right when there is decay, as compared to the no decay case.
The minimum and maximum values of $\sin^2\theta_{23}$ at 2$\sigma$s in the presence and absence of decay 
are shown in Table~\ref{s2tt23-pre-tab}. 

\begin{table}[htb]
% \centering
\begin{tabular}{|c|c|c|c|}
\hline
Analysis type & $\sin^2\theta_{23_{min}}(2\sigma)$ & $\sin^2\theta_{23_{max}}(2\sigma)$ & Precision at 1$\sigma$ (\%) \\
\hline
$\alpha_3=0$~eV$^2$ (fp) & 0.416 & 0.594 & 8.9\\
$\alpha_3=1\times10^{-5}$ eV$^2$ (fp) & 0.444 & 0.616 & 8.6\\
\hline
$\alpha_3=0$~eV$^2$ (marg) & 0.416 & 0.594 & 8.9\\
$\alpha_3=1\times10^{-5}$~eV$^2$ (marg) & 0.401 & 0.618 & 10.85\\
\hline
\end{tabular}
\caption{Minimum and maximum values of $\sin^2\theta_{23}$ at 2$\sigma$, with and without decay for fixed parameter and marginalised 
cases. The relative 1$\sigma$ precision obtained is also shown. NH is taken as the true hierarchy.}
\label{s2tt23-pre-tab}
\end{table}

\begin{figure}[htp]
\includegraphics[width=0.42\textwidth,height=0.40\textwidth]{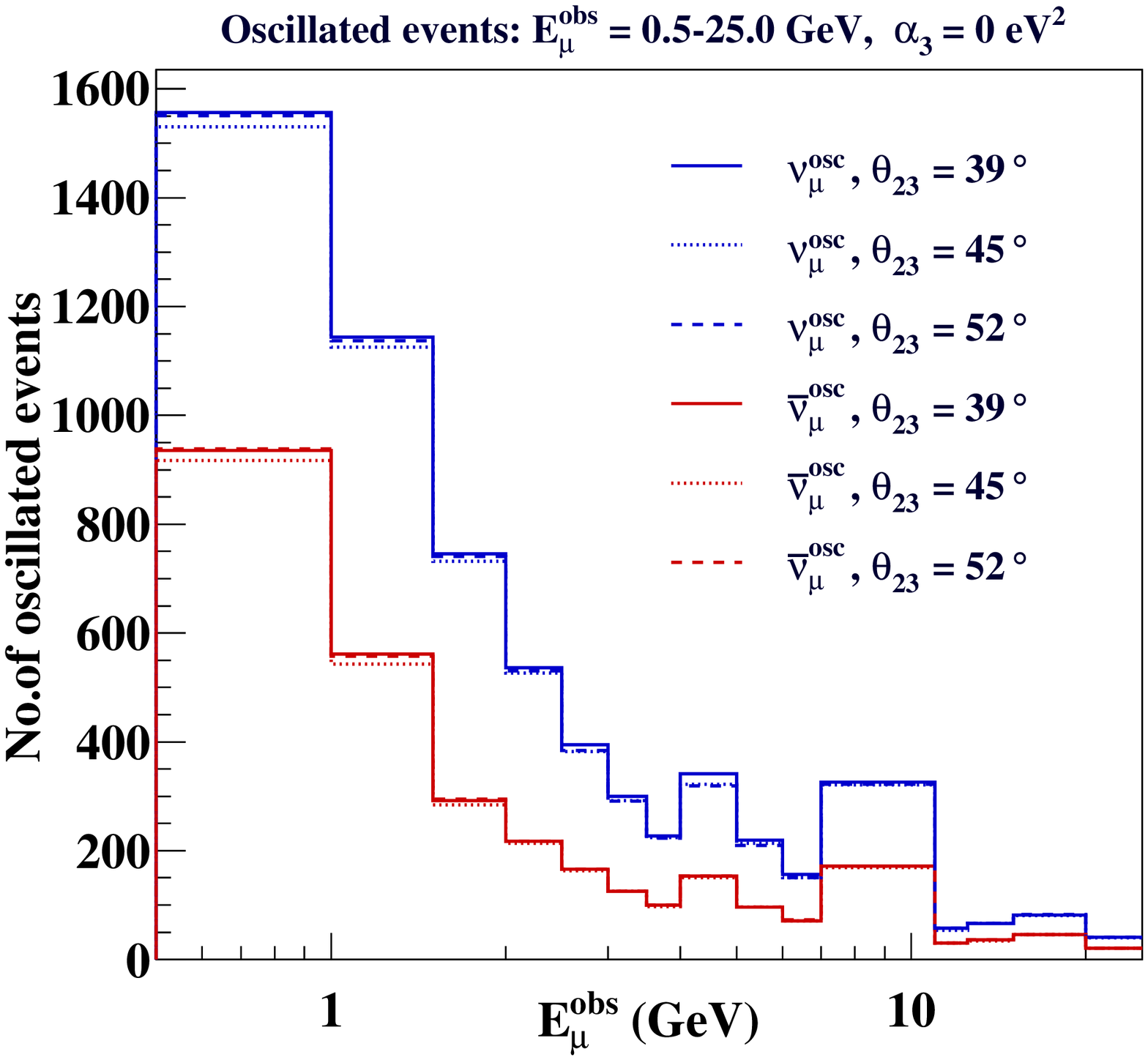}
\includegraphics[width=0.42\textwidth,height=0.40\textwidth]{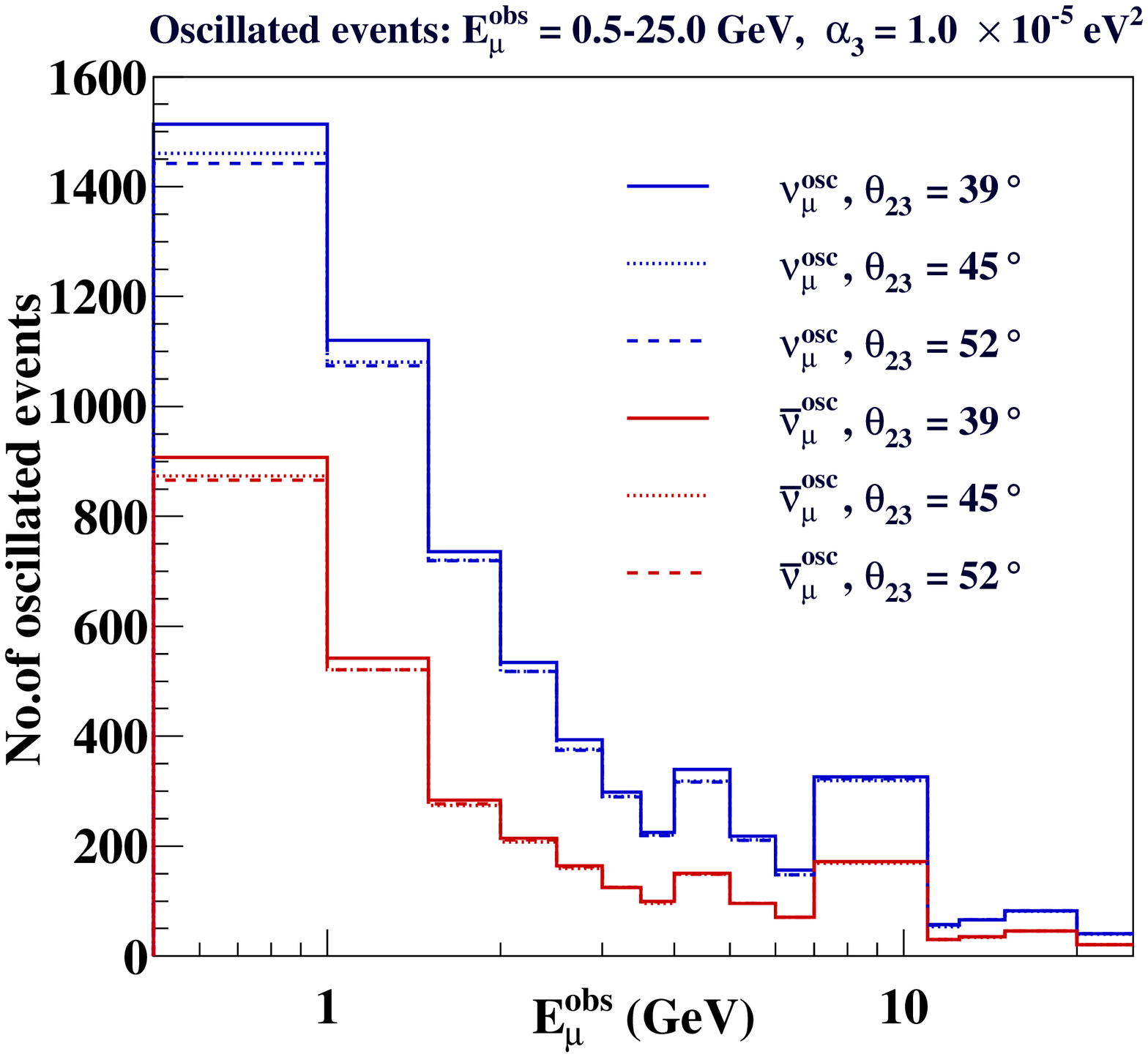}
\caption{Oscillated $\nu_\mu$ events as a function of $E_\mu$ for (left) $\alpha_3$ = 0~eV$^2$ 
and (right) $\alpha_3$ = $1\times10^{-5}$~eV$^2$ for $\theta_{23}$ = 39, 45 and 52$^\circ$.
}
\label{tt23-a3-nu-nubar}
\end{figure}

In order to understand the shift of parameter space, we show in 
Fig.~\ref{tt23-a3-nu-nubar}, the number of oscillated $\nu_\mu$ and 
$\bar{\nu}_\mu$ events for three different values of $\theta_{23}$ - 39$^\circ$, 45$^\circ$ and 
52$^\circ$ and two different $\alpha_3$ - 0 and $1\times10^{-5}$~eV$^2$. 
Here $39^\circ$ and $52^\circ$ are representative values for lower octant 
and higher octant respectively. We plot the events as a function of energy integrating over 
the zenith-angle bins. From the figures it can be seen that both  
in the absence and presence of decay there are differences between the 
number of events for various $\theta_{23}$ values. 
For the case of no decay this difference is less as compared to the case 
where decay is present. Comparing the figures on the left and right panels one also observes that, 
the difference between the  number of events for $\theta_{23}$ = 39$^\circ$ 
and 45$^\circ$ is more in presence of decay and the curve for 
$\theta_{23} = 45^\circ$ is closer to  $52^\circ$.  
Now, in obtaining the precision plot the data is generated 
with true $\theta_{23}$ = 45$^\circ$ and in theory the $\theta_{23}$ is 
kept fixed. For $\theta_{23}$ in the lower octant the difference of 
the number of events with that for $45^\circ$ being more in presence 
of decay, the $\chi^2$ for a $\theta_{23}$ 
in the lower octant will be higher as compared to the no decay case. 
On the other hand, for $\theta_{23}$ in the higher octant the difference 
in the number of events with $\theta_{23} = 45^\circ$ being less in presence 
of decay, one gets a lower $\chi^2$ as compared to the no decay case.
This explains why the precision curve shifts towards  higher $\theta_{23}$    
values. 

For the marginalised case, in the presence of decay the overall precision becomes worse compared to the no decay case. 
The 1$\sigma$ precision when decay is present is $\sim$10.85\% whereas for no decay it is $\sim$8.9\%. This can be explained as follows. 
In the marginalised case, for only oscillation we are trying to fit the ``data''
generated with $\theta_{23}$ = 45$^\circ$, varying the other parameters in theory. In this case the $\theta_{13}$ 
can be adjusted to give a slightly lower $\chi^2$. In presence of decay we generate the 
``data'' for a particular non-zero $\alpha_3$ and $\theta_{23}=45^\circ$. 
But now in theory we vary $\alpha_3$ as well as the other parameters. 
For $\theta_{23}$ in the lower octant, the theory events will be higher than the 
``data'' events as can be seen by comparing the events in the second panel of Fig.~\ref{tt23-a3-nu-nubar}.
However, in this case the $\alpha_3$ can be increased to give a better fit and a lower $\chi^2$. 
On the other hand for $\theta_{23}$ in the higher octant, the data events are higher than
the theory events and $\alpha_3$ can be decreased in theory to match 
the data better and give a lower $\chi^2$. This explains the widening of the $\chi^2$ vs $\theta_{23}$
curve in presence of decay. Note that this is more for the higher octant because the difference of 
the events for $\theta_{23}=45^\circ$ and say $52^\circ$ is less as compared to $\theta_{23}$ in the 
lower octant, say $39^\circ$. This gives a lower $\chi^2$ thus allowing more $\theta_{23}$ values in the 
higher octant.

\subsection{Precision on $|\Delta{m^2_{32}}|$ in the presence of decay}\label{pre-dm232-wd}
\begin{figure}[htp]
 \centering 
 \includegraphics[width=0.42\textwidth,height=0.40\textwidth]{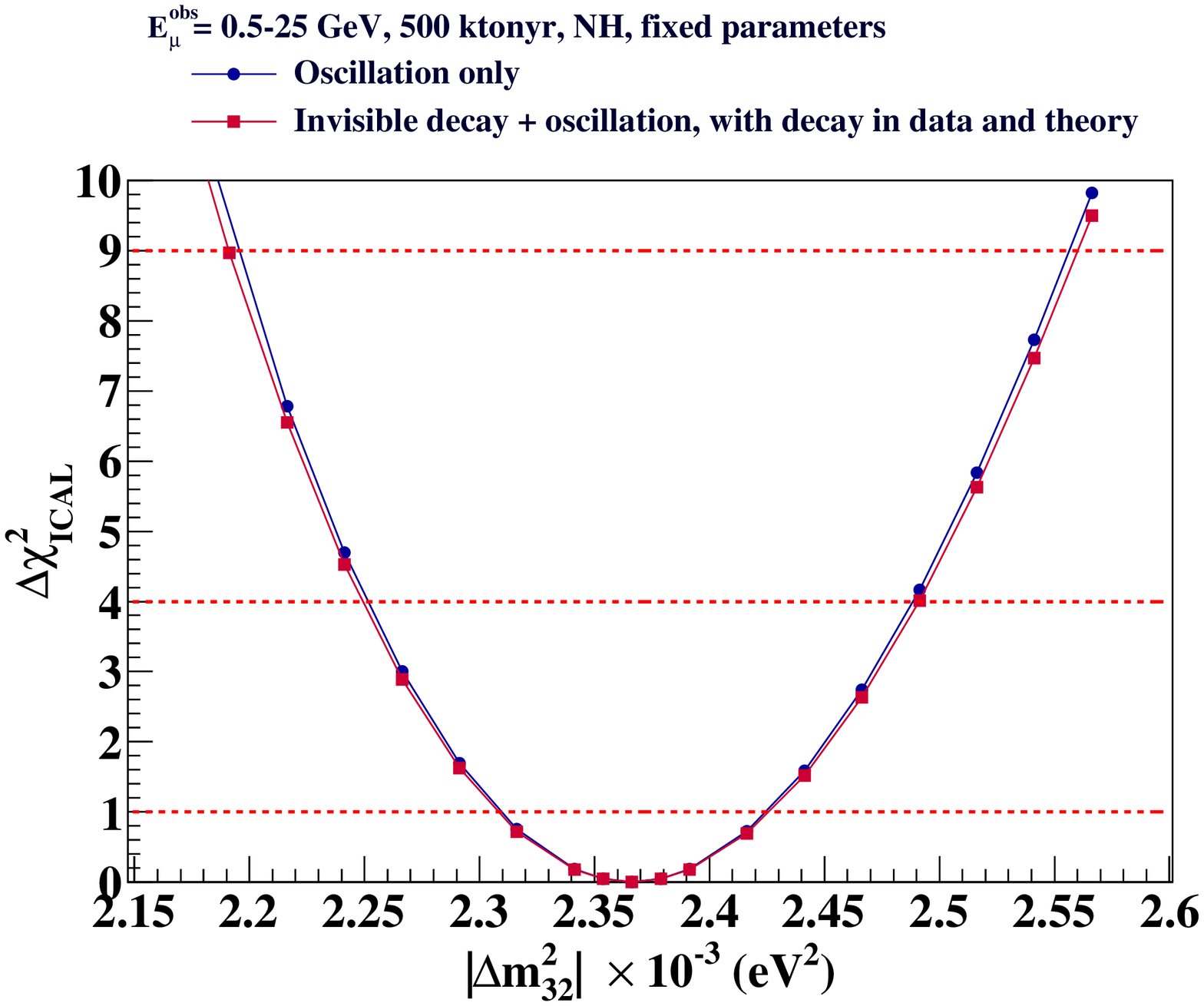}
  \includegraphics[width=0.42\textwidth,height=0.40\textwidth]{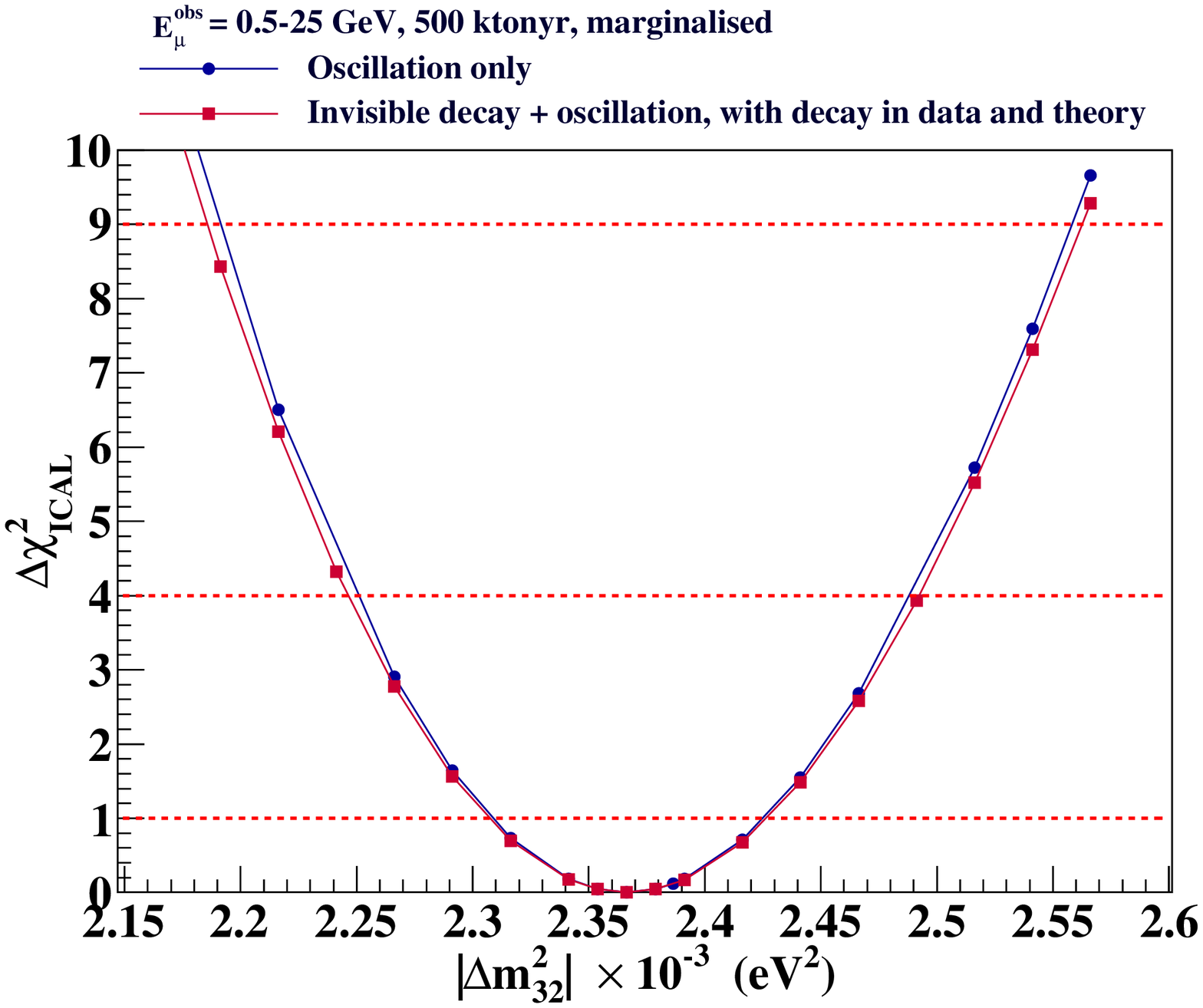}
 \caption{Precision on $|\Delta{m^2_{32}}|$ in the absence and presence of invisible decay 
 (left) fixed parameters (right) with marginalisation.}
 \label{dm232-pre-wd}
\end{figure}

:
The precision on the magnitude of the mass square difference $|\Delta{m^2_{32}}|$ in the presence and absence of 
   invisible decay of $\nu_3$ is presented in Fig.~\ref{dm232-pre-wd}. NH is taken
   as the true hierarchy. The relative 1$\sigma$ precision on $|\Delta{m^2_{32}}|$ with oscillations only  
   and with decay is $\sim$2.5\% for the fixed parameter case. When marginalisation is done this 
   becomes $\sim$2.6\% for both the cases. Thus it can be seen
   that the presence of decay does not affect the precision on $|\Delta{m^2_{32}}|$ much. This is because decay 
   mainly affects the amplitude of the oscillations and not the phase which is determined by $|\Delta{m^2_{32}}|$.  
   The minimum and maximum values of $\sin^2\theta_{23}$ at 2$\sigma$s in the presence and absence of decay 
   are shown in Table~\ref{dm232-pre-tab}. 
  
  \begin{table}[htb]
% \centering
\begin{tabular}{|c|c|c|c|}
\hline
Analysis type & $|\Delta{m^2_{32}}|_{min}(2\sigma)$ & $|\Delta{m^2_{32}}|_{max}(2\sigma)$ & Precision at 1$\sigma$ (\%) \\
	      & $\times10^{-3}\rm{eV}^2$            & $\times10^{-3}\rm{eV}^2$            &                      \\
\hline
$\alpha_3=0$~eV$^2$ (fp) & 2.252 & 2.489 & 2.5\\
$\alpha_3=1\times10^{-5}$~eV$^2$ (fp) & 2.249 & 2.492 & 2.5\\
\hline
$\alpha_3=0$~eV$^2$ (marg) & 2.252 & 2.489 & 2.6\\
$\alpha_3=1\times10^{-5}$~eV$^2$ (marg) & 2.247 & 2.493 & 2.6\\
\hline
\end{tabular}
\caption{Minimum and maximum values of $|\Delta{m^2_{32}}|$ at 2$\sigma$, with and without decay for fixed parameter and marginalised 
cases. The relative 1$\sigma$ precision obtained is also shown. NH is taken as the true hierarchy.}
\label{dm232-pre-tab}
\end{table}
% \clearpage
\subsection{Simultaneous precision on $\sin^2\theta_{23}$ and $|\Delta{m^2_{32}}|$
in the presence of $\alpha_3$}\label{contours} 
\begin{figure}[htp]
    \centering
     \includegraphics[width=0.5\textwidth,height=0.42\textwidth]{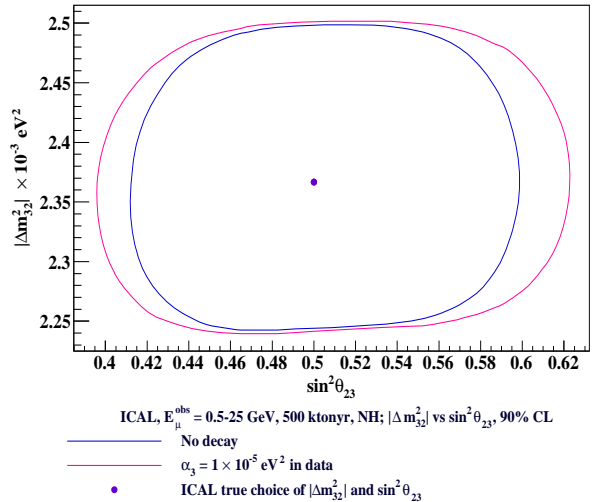}
     \caption{Expected 90\% C.L. contour in the $\sin^2\theta_{23}-|\Delta{m^2}_{32}|$ plane, with 
     and without decay, for NH. The value of $\alpha_3$ in ``data'' is taken as $1\times10^{-5}$~eV$^2$.}
     \label{tt23-dm232-4.61-pre}
    \end{figure}

  In this section the expected C.L. contours in the  
  $\sin^2\theta_{23}-|\Delta{m^2_{32}}|$ plane in the presence of decay are shown. The results are shown
  for true NH. A value of decay parameter $\alpha_3=1\times10^{-5}$~eV$^2$ is taken in ``data'' and is 
  marginalised in the 3$\sigma$ range $[0,2.35\times10^{-4}]$~eV$^2$. The other parameters are also 
  marginalised over their 3$\sigma$ ranges as before. The expected 90\% C.L. contour in the 
  $\sin^2\theta_{23}-|\Delta{m^2_{32}}|$ plane in the presence and absence of decay is shown in 
  Fig.~\ref{tt23-dm232-4.61-pre}.

 It can be seen that the precision worsens in the presence of decay. The contour widens significantly
along the $\sin^2\theta_{23}$ axis, more so in the second octant for the same reason as 
explained in the context of marginalised case in Fig.\ref{tt23-pre-wd}. In the absence of decay the precision 
on $\sin^2\theta_{23}$ at 90\% CL is 18.5\%. This worsens to 22.3\% with a decay parameter 
$\alpha_3=1\times10^{-5}$~eV$^2$. The precision on $|\Delta{m^2_{32}}|$ worsens only marginally from the 
no decay value of 5.35\% to 5.46\% for the same central value of $\alpha_3$. This is expected since the 
decay affects the oscillation amplitude which in turn affects the precision on $\sin^2\theta_{23}$.
% \clearpage   
\section{Summary and Discussions}\label{conclusion}
The expected sensitivity of ICAL to the decay lifetime of the mass eigenstate 
$\nu_3$, when it decays via the invisible decay mode was presented. The analysis was performed in the three-generation neutrino 
oscillation framework including decay as well as earth matter effects. 
The decay was parameterised in terms of $\alpha_3=m_3/\tau_3$, where, 
$m_3$ is the mass and $\tau_3$ the lifetime at rest of the mass eigenstate $\nu_3$. With 500 kton-yr of exposure, ICAL is expected to
constrain the invisible decay rate to  
$\alpha_3<4.36\times10^{-6}$~eV$^2$ at 90\% C.L., which is two orders of magnitude tighter than the 
bound obtained in \cite{gomes} for MINOS. In \cite{gomes} both charged current
(CC) and neutral current (NC) events were considered where as in our study only atmospheric CC $\nu_\mu$ and 
$\bar{\nu}_\mu$ events were used. For invisible neutrino decay, the NC background will be less. Hence 
the sensitivity to $\alpha_3$ is expected to improve.
%These studies involve the 
%invisible decay scenario only. However a recent study \cite{gago-gomes-vis} has shown that when the analysis
%is done by taking both visible as well as invisible decay into account, the bound on $\alpha_3$ improves 
%further. Hence in the case of ICAL also the bounds on $\alpha_3$ could improve if we study the visible
%decay + oscillation scenario also. So a combination of two decay scenarios and analysis of both CC and NC
%events should give a better bound on $\alpha_3$. This can be done in future. 

The effect of decay on the 2--3 oscillation parameters was also studied. Since the amplitude of oscillations is 
affected most by the presence of decay, it was found that decay affected the precision measurement of 
$\sin^2\theta_{23}$. For 500 kton-yrs of exposure assuming NH as the true hierarchy, 
the 1$\sigma$ precision on $\sin^2\theta_{23}$ was found to worsen to 10.85\% when $\alpha_3=10^{-5}$~eV$^2$ was
assumed. This is worse as compared to the 8.87\% obtained with oscillation only 
hypothesis. In the case of $|\Delta{m^2_{32}}|$ the 1$\sigma$ precision without decay is 2.5\%
whereas the inclusion of invisible decay does not affect it at all. The effect of $\alpha_3$ on the 
sensitivity to neutrino mass hierarchy and octant of $\theta_{23}$ will be studied elsewhere 
 \cite{hie-oct-decay}.

 It is also noteworthy that the sensitivity to smaller $\alpha_3$ comes mainly from the lower energy bins 
 below 2~GeV. Hence, if we can improve the efficiencies and resolutions of the detector, especially for muons 
 in the lower energy region, we will be able to put a better limit on $\alpha_3$. Reduction of the energy 
 threshold for the detection of low energy neutrinos in future will also help probing phenomena like decay with 
 increased precision. This is important since the atmospheric neutrino flux peaks at lower energies and by being 
 able to detect and analyse more events we will further improve our sensitivities to all parameters including $\alpha_3$.

% We further want to point out that the bounds obtained in this paper are using atmospheric neutrinos 
% only. Our analysis shows that even with atmospheric neutrinos where the fluxes cannot be controlled, 
% if we have a huge magnetised detector which enables the separation of $\nu_\mu$ and $\bar{\nu}_\mu$,
% that itself can give good sensitivity to $\alpha_3$. The further addition of NC events, visible
% decay scenario, improvement of the efficiencies and resolutions etc as mentioned previously will help
% in putting a better bound on $\alpha_3$. These results combined with the results from accelerator 
% and reactor experiments can give a tighter limit on the decay parameter. 

 \section*{Acknowledgments}
 The authors acknowledge Prof.~Amol~Dighe, TIFR, Mumbai and Prof.~D.~Indumathi IMSc, Chennai for the
 discussions. We also acknowledge our INO internal referees, Prof.~S.~Uma Sankar, IITB, Mumbai 
 and Prof.~Jim~Libby, IITM, Chennai for their comments and suggestions. 
 LSM acknowledges Nandadevi and Satpura clusters belonging to the computing facility of IMSc Chennai, 
 with which this analysis was made possible.

\end{document}